\documentclass[a4paper,12pt]{article}
\usepackage{amssymb}
\usepackage{amsmath}
\usepackage{latexsym,amssymb,amsmath}
\usepackage{graphicx}

\DeclareFontFamily{U}{rsf}{} \DeclareFontShape{U}{rsf}{m}{n}{
  <5> <6> rsfs5 <7> <8> <9> rsfs7 <10-> rsfs10}{}
\DeclareMathAlphabet\Scr{U}{rsf}{m}{n} \makeatletter
\@addtoreset{equation}{section} \makeatother

\newcommand{\be}{\begin{equation}}
\newcommand{\ee}{\end{equation}}
\newcommand{\bea}{\begin{eqnarray}}
\newcommand{\eea}{\end{eqnarray}}
\newcommand{\ba}{\begin{array}}
\newcommand{\ea}{\end{array}}
\newcommand{\bit}{\begin{itemize}}
\newcommand{\eit}{\end{itemize}}
\newcommand{\ben}{\begin{enumerate}}
\newcommand{\een}{\end{enumerate}}

\begin{document}

\begin{titlepage}
\begin{center}
\hfill CERN-PH-TH/2010-301 \\
\hfill SU-ITP-10/09 \\
\hfill KCL-MTH-10-12 \\

\vskip 1.5cm

{\Huge \bf  Exceptional Reductions}

\vskip 1.5cm

{\bf Alessio Marrani\,$^1$, Emanuele Orazi\,$^1$ and Fabio
Riccioni\,$^2$}

\vskip 20pt

{\it ${}^1$ Physics Department, Theory Unit, CERN,\\
     CH -1211, Geneva 23, Switzerland\\\vskip 5pt
     \texttt{Alessio.Marrani@cern.ch}\\
     \texttt{orazi@lnf.infn.it}}

     \vspace{10pt}

{\it $^2$ Department of Mathematics, King's College London \\
Strand London WC2R 2LS UK\\\vskip 5pt
\texttt{Fabio.Riccioni@kcl.ac.uk}}

\end{center}

\vskip 2.2cm

\begin{center} {\bf ABSTRACT}\\[3ex]\end{center}
Starting from basic identities of the group $E_{8}$, we perform
progressive reductions, namely decompositions with respect to the
maximal and symmetric embeddings of $E_{7}\times SU\left( 2\right) $
and then of $E_{6}\times U\left( 1\right) $. This procedure provides
a systematic approach to the basic identities involving invariant
primitive tensor structures of various irreprs. of
finite-dimensional exceptional Lie groups. We derive novel
identities for $E_{7}$ and $E_{6}$, highlighting the $E_{8}$ origin
of some well known ones. In order to elucidate the connections of
this formalism to four-dimensional Maxwell-Einstein supergravity
theories based on symmetric scalar manifolds (and related to
irreducible Euclidean Jordan algebras, the unique exception being
the triality-symmetric $\mathcal{N}=2$ $stu$ model), we then derive
a fundamental identity involving the unique rank-$4$ symmetric
invariant tensor of the $0$-brane charge symplectic irrepr. of
$U$-duality groups, with potential applications in the quantization
of the charge orbits of supergravity theories, as well as in the
study of multi-center black hole solutions therein.




\vfill


\end{titlepage}

\newpage \setcounter{page}{1} \numberwithin{equation}{section}

\newpage\tableofcontents

\section{Introduction}

Supergravity theories have a rich algebraic structure, which also
reflects into their scalar manifolds. A particularly remarkable
class of scalar manifolds is given by the homogeneous spaces $G/H$,
with $G$ a non-compact Lie group, and $H$ denoting its maximal
compact subgroup. In particular, in maximal supergravities the $U$-duality\footnote{%
Here $U$-duality is referred to as the ``continuous'' symmetries of
\cite {CJ-1}. Their discrete versions are the $U$-duality
non-perturbative string theory symmetries introduced by Hull and
Townsend \cite{HT-1}.} groups $G$ belong to the so-called
exceptional $E_{n\left( n\right) }$-sequence \cite {J-1,J-2} of
symmetries of theories in $11-n$ dimensions. This sequence is
encoded in the very-extended Kac-Moody algebra $E_{11}$ \cite{west},
and each theory corresponds to a decomposition with respect to each
subalgebra $GL(11-n,\mathbb{R})\times E_{n(n)}$. When $n=9$, that is
in two dimensions, the field equations of the theory possess an
$E_{9(9)}$ symmetry, which is the infinite dimensional affine
extension of $E_{8(8)}$ \cite{J-2,nicolai}. This is completely
general: the (on-shell) symmetry of a two-dimensional theory
obtained from the reduction of a three-dimensional theory whose
scalars parametrise the manifold $G/H$ is the infinite dimensional
affine extension of $G$.

Since the process of dimensional reduction leads to infinite
dimensional symmetries in $D\leqslant 2$, when one confines his
attention to finite-dimensional symmetry groups the endpoint of a
chain of symmetries of theories related by dimensional reduction is
$D=3$. As shown in \cite{14,15}, and further systematized in \cite
{Ox-1,Ox-2} (elaborating on ideas and results on ``group
disintegrations'' of \cite{J-1,7-2}; see also \cite{solv}), one has
a group-theoretic framework to determine which $D=3$ theories can be
conceived as dimensional reductions of higher-dimensional theories.
In particular, in \cite{Ox-2} the systematics of oxidations
involving \textit{non-split} $U$-dualities, including the bosonic
sectors of the theories with $8$ supersymmetries based on symmetric
scalar manifolds, related by the $r$- and $c$- maps
\cite{CFG-1,11-2,dWVVP}, has been developed using the diagrammatic
language of Tits-Satake diagrams (see \textit{e.g.} \cite {TS-rev},
and Refs. therein). This result was systematised in \cite{RVPW},
where it was shown that starting from the Tits-Satake diagram of the
three-dimensional theory one can construct a very-extended Kac-Moody
algebra such that its Dynkin diagram encodes all the properties of
the theory in various dimensions. The reconstruction procedure,
which allows to determine the higher-dimensional ancestor(s) of a
lower-dimensional theory, is usually named \textit{``oxidation''}. A
remarkable aspect of oxidation is that, differently from the
dimensional reduction, it is not unique, in the sense that it can
admit different ``branches'', namely distinct higher-dimensional
theories (eventually related by string dualities) originating the
same lower-dimensional theory upon dimensional reduction.

The investigation presented in this paper approaches the oxidations
from the point of view of the fundamental identities involving
invariant primitive tensor structures of the relevant (namely,
fundamental and adjoint) irreprs. of the $U$-duality groups.
Confining ourselves to finite-dimensional groups, we start from
basic identities in the adjoint irrepr. $\mathbf{248}$ of $E_{8}$,
and we perform progressive reductions, given by decompositions with
respect to the maximal and symmetric $E_{8}$-embedding of
$E_{7}\times SU\left( 2\right) $ and then to the maximal and
symmetric $E_{7}$-embedding of $E_{6}\times U\left( 1\right) $.
Within such a framework, this approach provides a systematic way to
derive all basic identities describing the structure of $E_{8}$,
$E_{7}$ and $E_{6}$ exceptional Lie groups. Indeed, we derive many
novel identities involving the relevant invariant primitive tensors
of such groups, and we also highlight the common origin (through
iterated reduction) of some well known identities. Furthermore, we
also present some results on the further maximal and symmetric
$E_{6}$-embedding of $SO\left( 10\right) \times U\left( 1\right) $,
retrieving various well known Fierz identities of $SO\left(
10\right) $.

\begin{table}[p]
\begin{center}
\begin{tabular}{|c||c|c|c|}
\hline $
\begin{array}{c}
\\
J_{3}
\end{array}
$ & $
\begin{array}{c}
\\
G_{5} \\
~~
\end{array}
$ & $
\begin{array}{c}
\\
G_{4} \\
~~
\end{array}
$ & $
\begin{array}{c}
\\
G_{3} \\
~~
\end{array}
$ \\ \hline\hline $
\begin{array}{c}
\\
J_{3}^{\mathbb{O}_{s}} \\
~
\end{array}
$ & $E_{6\left( 6\right) }~$ & $
\begin{array}{c}
E_{7\left( 7\right) } \\
d=133,f=56
\end{array}
$ & $E_{8\left( 8\right) }~$ \\ \hline $
\begin{array}{c}
\\
J_{3}^{\mathbb{O}} \\
~
\end{array}
$ & $E_{6\left( -26\right) }~$ & $
\begin{array}{c}
E_{7\left( -25\right) } \\
d=133,f=56
\end{array}
$ & $E_{8\left( -24\right) }$ \\ \hline $
\begin{array}{c}
\\
J_{3}^{\mathbb{H}} \\
~
\end{array}
$ & $SU^{\ast }\left( 6\right) $ & $
\begin{array}{c}
SO^{\ast }\left( 12\right) \\
d=66,f=32
\end{array}
$ & $E_{7\left( -5\right) }$ \\ \hline $
\begin{array}{c}
\\
J_{3}^{\mathbb{C}} \\
~
\end{array}
$ & $SL\left( 3,\mathbb{C}\right) $ & $
\begin{array}{c}
SU\left( 3,3\right) \\
d=35,f=20
\end{array}
~$ & $E_{6\left( 2\right) }$ \\ \hline $
\begin{array}{c}
\\
J_{3}^{\mathbb{R}} \\
~
\end{array}
$ & $SL\left( 3,\mathbb{R}\right) $ & $
\begin{array}{c}
Sp\left( 6,\mathbb{R}\right) \\
d=21,f=14
\end{array}
$ & $F_{4\left( 4\right) }~$ \\ \hline $
\begin{array}{c}
\\
M_{1,2}\left( \mathbb{O}\right) \\
~
\end{array}
$ & $-$ & $
\begin{array}{c}
SU\left( 1,5\right) \\
d=35,f=20
\end{array}
$ & $E_{6\left( -14\right) }$ \\ \hline $
\begin{array}{c}
\\
\mathbb{R} \\
~
\end{array}
$ & $-$ & $
\begin{array}{c}
SL\left( 2,\mathbb{R}\right) \\
d=3,f=4
\end{array}
$ & $G_{2\left( 2\right) }$ \\ \hline $
\begin{array}{c}
\\
\mathbb{R}\oplus \mathbb{R}\oplus \mathbb{R} \\
~
\end{array}
$ & $\left[ SO\left( 1,1\right) \right] ^{2}$ & $
\begin{array}{c}
\left[ SL\left( 2,\mathbb{R}\right) \right] ^{3} \\
d=9,f=8
\end{array}
$ & $SO\left( 4,4\right) $ \\ \hline
\end{tabular}
\end{center}
\caption{$U$-duality groups $G_{3}$, $G_{4}$ and $G_{5}$ related to
\textit{irreducible} rank-$3$ Euclidean Jordan algebras in (Minkowskian) $D=3$, $4$ and $5$%
, respectively. Also $d\equiv $dim$_{\mathbb{R}}\mathbf{Adj}\left(
G_{4}\right) $ and $f\equiv $dim$_{\mathbb{R}}\mathbf{R}\left(
G_{4}\right) $ are given. The corresponding scalar manifolds are the \textit{%
symmetric} cosets $\frac{G}{H}$, where $H$ is the maximal compact
subgroup (with symmetric embedding) of $G$. $\mathbb{O}$,
$\mathbb{H}$, $\mathbb{C}$ and $\mathbb{R}$ respectively denote the
four division algebras of octonions, quaternions, complex and real
numbers, and $\mathbb{O}_{s}$ is the split form of octonions.
$M_{1,2}\left( \mathbb{O}\right) $ is the Jordan triple system (not
upliftable to $D=5$) generated by $2\times 1$ matrices over
$\mathbb{O}$ \protect\cite{GST}. Note that the $stu$ model, based on
$\mathbb{R}\oplus \mathbb{R}\oplus \mathbb{R}$, is reducible, but
\textit{triality symmetric}. All cases pertain to models with $8$
supersymmetries, with exception of $M_{1,2}\left( \mathbb{O}\right) $ and $%
J_{3}^{\mathbb{O}_{s}}$, related to $20$ and $32$ supersymmetries,
respectively. The $d=5$ uplift of the $t^{3}$ model based on
$\mathbb{R}$ is the \textit{pure} $\mathcal{N}=2$, $D=5$
supergravity. $J_{3}^{\mathbb{H}}$ is related to both $8$ and $24$
supersymmetries, because the corresponding supergravity theories
share the very same bosonic sector \protect\cite
{GST,ADF-U-duality-d=4,Samtleben-Twin}. Note that the dimensions $f$
and $d$ of \textit{all} reported $G_{4}$'s satisfy the relation
$d=\frac{3f\left( f+1\right) }{f+16}. $  In particular, by
considering $\mathcal{N}=2$, $D=4$ ``magic''
supergravities (based on $J_{3}^{\mathbb{A}}$) as well as $\mathcal{N}=8$, $%
D=4$ maximal supergravity (based on $J_{3}^{\mathbb{O}_{s}}$), and
defining $q\equiv \text{dim}_{\mathbb{R}}\mathbb{A}=8,4,2,1\text{~for~}\mathbb{A}=%
\mathbb{O},\mathbb{H},\mathbb{C},\mathbb{R}$, the relation $f=6q+8$
yields to $d=\frac{3\left( 3q+4\right) \left( 2q+3\right) }{q+4}.$
Note that these relations also admit a limit $q=0$, reproducing $d$
and $f$ of the reducible yet \textit{triality-symmetric} $stu $
model.}
\end{table}

Our procedure applies to the $D=3\rightarrow 4\rightarrow
5\rightarrow 6$ entries of the $E_{n\left( n\right) }$ exceptional
sequence of Cremmer-Julia, pertaining to split forms and thus to
maximal supergravity in $D=3,4,5,6$, related to the irreducible
Euclidean Jordan algebra over the split form of the octonions
$\mathbb{O}_{s}$:
\begin{equation}
E_{8\left( 8\right) }\longrightarrow E_{7\left( 7\right)
}\longrightarrow E_{6\left( 6\right) }\longrightarrow SO\left(
5,5\right) ; \label{CJ-seq-(1)}
\end{equation}
in particular, $SO\left( 5,5\right) \equiv E_{5\left( 5\right) }$ is the $U$%
-duality group of the $\left( 2,2\right) $ non-chiral maximal $D=6$
supergravity based on\footnote{$\mathbf{\Gamma }_{m,n}$ stands for
the
Jordan algebra of degree two with a quadratic form of Lorentzian signature $%
\left( 5,5\right) $, which is nothing but the Clifford algebra of
$O\left( 5,5\right) $ \cite{JVNW}.
\par
Also note the maximal (symmetric) algebraic embedding
\begin{equation*}
J_{3}^{\mathbb{O}_{s}}\supset ^{\max }\mathbb{R}\oplus J_{2}^{\mathbb{O}%
_{s}}.
\end{equation*}
} $J_{2}^{\mathbb{O}_{s}}\sim \mathbf{\Gamma }_{5,5}$. However,
since we do not specify the non-compact real form of the groups
under consideration, our procedure also applies to the following
non-split version of the
Cremmer-Julia sequence (\ref{CJ-seq-(1)})\footnote{%
In the theories with $8$ supersymmetries, the so-called $c$-map
\cite{CFG-1} and $r$-map (see \textit{e.g.} \cite{dWVVP,LA08-Proc}
for more Tables and a list of Refs.) relate $D=4/D=3$ and $D=5/D=4$,
respectively.}:
\begin{equation}
E_{8\left( -24\right) }\overset{c}{\longrightarrow }E_{7\left( -25\right) }%
\overset{r}{\longrightarrow }E_{6\left( -26\right) }\longrightarrow
SO\left( 1,9\right) ;  \label{non-split-CJ-seq-1}
\end{equation}
in particular, $SO\left( 1,9\right) $ is the $U$-duality group of the $%
\left( 1,0\right) $ chiral minimal $D=6$ magic supergravity based on%
\footnote{%
Note the maximal (symmetric) algebraic embedding (see \textit{e.g.}
\cite {Gun-Refs})
\begin{equation*}
J_{3}^{\mathbb{A}}\supset ^{\max }\mathbb{R}\oplus
J_{2}^{\mathbb{A}},
\end{equation*}
where $J_{2}^{\mathbb{A}}\sim \mathbf{\Gamma }_{1,q+1}$, with $q\equiv $dim$%
_{\mathbb{R}}\left( \mathbb{A}\right) =8,4,2,1$ for the division algebras $%
\mathbb{A}=\mathbb{O},\mathbb{H},\mathbb{C},\mathbb{R}$, respectively.} $%
J_{2}^{\mathbb{O}}\sim \mathbf{\Gamma }_{1,9}$. This sequence has an
interpretation in terms of iterated oxidations pertaining to the
``magic'' supergravity theory with $8$ supersymmetry in
$D=3\rightarrow 4\rightarrow 5\rightarrow 6$, related to the
irreducible Euclidean Jordan algebra over the division algebra of
the octonions $\mathbb{O}$.

Remarkably, the results on the fundamental identities of $E_{8}$,
$E_{7}$ and $E_{6}$ also hold for the first three elements of the
sequence
\begin{equation}
\underset{J_{3}^{\mathbb{O}}}{E_{8\left( -24\right)
}}\longrightarrow \underset{J_{3}^{\mathbb{H}}}{E_{7\left( -5\right)
}}\longrightarrow \underset{J_{3}^{\mathbb{C}}}{E_{6\left( 2\right)
}}\longrightarrow \underset{J_{3}^{\mathbb{R}}}{F_{4\left( 4\right)
}}\longrightarrow
\underset{\mathbb{R}\oplus \mathbb{R}\oplus \mathbb{R}}{SO\left( 4,4\right) }%
\longrightarrow \underset{\mathbb{R}}{G_{2\left( 2\right) }}.
\label{new?-no-MS-1}
\end{equation}
This sequence has not an interpretation in terms of iterated
oxidation, but, as reported in Table 1, it rather describes a chain
of embeddings of the $D=3 $ $U$-duality groups of supergravity
theories with symmetric scalar manifolds associated to
\textit{irreducible} Euclidean Jordan algebras,
given in the second line of (\ref{new?-no-MS-1}). The cases $J_{3}^{\mathbb{A%
}}$ ($\mathbb{A}=\mathbb{O},\mathbb{H},\mathbb{C}$ and $\mathbb{R}$
being the four division algebras of octonions, quaternions, complex
and real numbers) correspond to ``magic'' supergravities \cite{GST},
whereas the cases $\mathbb{R}\oplus \mathbb{R}\oplus \mathbb{R}$ and
$\mathbb{R}$
respectively pertain to the $c$-map of the so-called $\mathcal{N}=2$, $D=4$ $%
stu$ \cite{stu} and $t^{3}$ models. The corresponding cosets can all
be obtained by a $c$-map \cite{CFG-1} of suitable symmetric special
K\"{a}hler
manifolds. Clearly, the rank-$3$ Jordan algebra $\mathbb{R}\oplus \mathbb{R}%
\oplus \mathbb{R}$ is not irreducible; however, it is \textit{sui
generis}, because it enjoys the remarkable \textit{triality
symmetry} \cite{stu}.

Also note that the sequence (\ref{CJ-seq-(1)}) is related to the sequence (%
\ref{new?-no-MS-1}) by the maximal symmetric embedding $E_{8\left(
8\right) }\supset E_{7\left( -5\right) }$, which has the trivial
supergravity interpretation that $\mathcal{N}=4$, $d=3$
$J_{3}^{\mathbb{H}}$-magic theory is a consistent truncation of
maximal $\mathcal{N}=16$, $d=3$ supergravity.

It is worth remarking that in our treatment we will not restrict to
dimensional reductions on purely spacelike internal manifolds
(usually \textit{tori}). As resulting from the analyses of
\cite{14,5-3}, the only group theoretical difference between
timelike and spacelike reductions is the non-compact nature of the
coset stabilizer $H$. Recently, timelike reductions to $D=3$ have
been used as an efficient tool to describe and classify spherically
symmetric, asymptotically flat and stationary black hole solutions
(and the corresponding scalar flows) of $D=4$ supergravity theories
with symmetric scalar manifolds (see \textit{e.g.} \cite{Trig-1},
and Refs. therein). Interestingly, this also turned out to be
relevant
within the so-called ``black hole/qubit correspondence'' \cite{BH/qubit-Refs}%
.\medskip

In general, our group-theoretical approach to oxidation can be
considered as complementary to the one exploited in
\cite{J-1,7-2,14,15,Ox-1,Ox-2}, because we deal with the reductions
of the identities involving the invariant primitive tensors of the
relevant irreprs. of the $U$-duality groups. This procedure provides
a systematic derivation of a number of fundamental identities
characterizing the $U$-duality groups of supergravity theories in
various dimensions.\smallskip

As application of the results on the oxidations of group structure
identities discussed above, we will then derive an identity
involving the so-called $K$\textit{-tensor}, namely the unique
rank-$4$ symmetric invariant tensor of the irrepr. of $G_{4}$ in
which the black hole charges sit. When contracted with four charge
vectors, the $K$-tensor gives rise to the $G_{4}$-invariant
homogeneous quartic polynomial $I_{4}$, which plays a prominent role
in the algebraic classification of the charge orbits and ``moduli
spaces'' of extremal black hole attractors \cite{AM} in $d=4$
Maxwell-Einstein supergravities (see \textit{e.g.}
\cite{ADFT-review}-\cite {N=2-D=4-Dressed}, and Refs. therein).
Moreover, the identity we will derive has potential applications in
\textit{at least} two other frameworks, namely: \textit{i}) the
quantization of the charge orbits of supergravity theories, which
might be relevant in relation to recent developments on the possible
$UV$-finiteness of $\mathcal{N}=8$, $D=4$ supergravity (see
\textit{e.g.} \cite{N=8-UV-finiteness}, and Refs. therein);
\textit{ii}) the group-theoretical study of $U$-invariants relevant
for multi-center black holes \cite{FMOSY-1,irred-1}.
\bigskip

The plan of the paper is as follows. In Sect. \ref{Sect-2} we
introduce the notation, and we report some general results on the
relation between various data characterizing some Lie groups,
appearing, through suitable non-compact real forms, as $U$-duality
groups of supergravity theories. We also consider identities
involving up to four structure constants, holding true for all
finite-dimensional exceptional Lie groups. Sects. \ref{E8--->E7} and
\ref{E7--->E6} exploit the approach based on the progressive
oxidation of the starting $E_{8}$-identities involving up to four
structure constants. As discussed above, this amounts to decomposing
such identities with respect to the following chain of maximal and
symmetric group embeddings:
\begin{equation}
E_{8}\supset E_{7}\times SU(2)\supset E_{6}\times SU(2)\times
U\left( 1\right) ,
\end{equation}
and it provides a systematic way to derive all $E_{7}$-identities and $E_{6}$%
-identities originating from the basic starting relations for
$E_{8}$. We name this method \textit{``exceptional reductions''}.
In Sect. \ref{K-Tensor} we derive a fundamental identity involving the $K$%
-tensor. Besides the aforementioned importance of the $K$-tensor for
the theory of extremal black hole attractors \cite {AM} in
Maxwell-Einstein supergravities (see Sec. \ref{K-Tensor}), this
(hitherto unknown)
result has potential application in the issue of the classification of the orbits of the irrepr. $%
\mathbf{R}\left( G_{4}\right) $ in presence of
Dirac-Zwanziger-Schwinger charge quantization conditions, especially
for $\mathcal{N}=8$, $D=4$ supergravity (see \textit{\ e.g.} \cite
{Krut-1,Duff-1,Duff-2,N=8-UV-finiteness}, and Refs. therein), as
well as in the study of multi-center black holes \cite
{FMOSY-1,irred-1}.

Various details and further results are given in the three
Appendices which conclude the paper. In App. \ref{SU(2)-Convs} we
summarise our conventions for $SU(2)$, crucial in order to perform
the reduction of $E_{8}$-identities in Sect. \ref{E8--->E7}. In App.
\ref{SO(10)} we further reduce some $E_{6}$-identities obtained in \
Sect. \ref{E7--->E6} with respect to the maximal and symmetric
embedding
\begin{equation}
E_{6}\supset SO\left( 10\right) \times U\left( 1\right)
,\label{E6-emb}
\end{equation}
retrieving some well known $SO\left( 10\right) $ Fierz identities,
whose common origin (through iterated reduction) is thus clarified.
In App. \ref {Decomps} we derive an useful group theoretical
decomposition used in Sect. \ref{K-Tensor}, holding \textit{at
least} for all $G_{4}$'s reported in Table 1.

\section{\label{Sect-2}Preliminaries}

The present Section is aimed at introducing the notation used
throughout the paper, and at discussing the general approach which
we will follow. Furthermore, some basic identities for the
exceptional Lie group $E_{8}$ will be derived, which will then be
used in the analysis of Sects. \ref {E8--->E7} and \ref{E7--->E6},
in turn leading to other basic identities for the exceptional groups
$E_{7}$ and $E_{6}$, respectively.

For a generic simple Lie group $G$ the Cartan-Killing metric $\kappa
^{\alpha \beta }$ is defined as
\begin{equation}
C_{\mathbf{Adj}}\kappa ^{\alpha \beta }=f^{\alpha \gamma
}{}_{\epsilon }f^{\beta \epsilon }{}_{\gamma },
\label{standardcartankilling}
\end{equation}
where $C_{\mathbf{Adj}}$ is the quadratic Casimir in the adjoint irrepr. $%
\mathbf{Adj}$ (with lowercase Greek indices), and $f^{\alpha \beta
}{}_{\gamma }$ are the structure constants of the corresponding Lie algebra $%
\frak{g}$.

We then consider the charge irrepr. $\mathbf{R}$ of the $U$-duality group $G$%
. In $D=4$, $\mathbf{R}=\mathbf{Sympl}$, namely it is the smallest
non-trivial symplectic irrepr. of $G$ (\textit{e.g.} $\mathbf{R}=\mathbf{Fund%
}=\mathbf{56}$ for $E_{7}$) , with a unique singlet
$\mathbb{C}_{MN}$ (the
symplectic metric) in its antisymmetric tensor product\footnote{%
The subscripts ``$s$'' and ``$a$'' respectively denote the symmetric
and antisymmetric tensor products throughout.}:
\begin{equation}
\exists !\mathbb{C}_{MN}\equiv \mathbf{1}\in \mathbf{Sympl}_{a}^{2}.
\label{sympl-def}
\end{equation}
In $D=5$, $\mathbf{R}$ is not symplectic, but rather it splits into
two
(electric and magnetic) charge irreprs. (\textit{e.g.} for $E_{6}$: $\mathbf{%
R}=\mathbf{Fund}=\mathbf{27}$ gradient, and $\overline{\mathbf{R}}=\overline{%
\mathbf{Fund}}=\overline{\mathbf{27}}$ contragradient). The $D=3$ case of $%
E_{8}$ stands on its own, because $\mathbf{R}=\mathbf{Fund}=\mathbf{Adj}=%
\mathbf{248}$; see treatment of Sects. \ref{E8--->E7},
\ref{E7--->E6} and \ref{K-Tensor} for further elucidation.

The quadratic Casimir $C_{\mathbf{R}}$ of $\mathbf{R}$ (with
uppercase Latin indices) is defined \textit{via}
\begin{equation}
C_{\mathbf{R}}\delta _{M}^{N}=\kappa _{\alpha \beta }t_{M}^{\alpha
}{}^{P}t_{P}^{\beta }{}^{N},  \label{casimirfund}
\end{equation}
where $t_{M}^{\alpha }{}^{N}$ are the generators of $\frak{g}$ in $\mathbf{R}%
\left( G\right) $:
\begin{equation}
\left[ t^{\alpha },t^{\beta }\right] _{M}^{~N}=f^{\alpha \beta
}{}_{\gamma }t_{M}^{\gamma }{}^{N}.  \label{tt=ft}
\end{equation}

In this paper we will adopt instead a different metric (see for
instance the appendix of \cite{Trig-2}), namely
\begin{equation}
g^{\alpha \beta }=Tr(t^{\alpha }t^{\beta })=t_{M}^{\alpha
}{}^{N}t_{N}^{\beta }{}^{M}.  \label{ourmetric}
\end{equation}
This in turn implies
\begin{equation}
g_{\alpha \beta }t_{M}^{\alpha }{}^{P}t_{P}^{\beta
}{}^{N}=\frac{d}{f}\delta _{M}^{N},  \label{gDDisdoverdlambda}
\end{equation}
where $d\equiv $dim$_{\mathbb{R}}\mathbf{Adj}\left( G\right) $ and $f\equiv $%
dim$_{\mathbb{R}}\mathbf{R}\left( G\right) $ throughout.

Within this notation, Eq. (\ref{standardcartankilling}) is replaced
by (see \textit{e.g.} \cite{Trig-2,Riccioni-Hierarchy} and Refs.
therein)
\begin{equation}
f_{\alpha \gamma \delta }f_{\beta }^{~\gamma \delta }=-\frac{d}{f}{\frac{C_{%
\mathbf{Adj}}}{C_{\mathbf{R}}}}g_{\alpha \beta }=-{\frac{g^{\vee }}{%
\widetilde{I}}}g_{\alpha \beta },  \label{ff=deltacasimir}
\end{equation}
where $g^{\vee }$ is the dual Coxeter number of $G$, and
$\widetilde{I}$ is the Dynkin index of $\mathbf{R}\left( G\right)
$.\medskip

The results obtained in the present paper, and in particular in Sec.
\ref
{K-Tensor}, hold \textit{at least} for all $D=4$ $U$-duality groups $%
G_{4}$'s related to \textit{irreducible} rank-$3$ Euclidean Jordan
algebras,
with the only exception of $stu$ model (related to the \textit{%
triality-symmetric}, \textit{reducible} rank-$3$ Jordan algebra $\mathbb{R}%
\oplus \mathbb{R}\oplus \mathbb{R}$). Such groups are reported in
Table 1,
along with their corresponding $D=3$ and $D=5$ counterparts\footnote{%
The trivial Jordan algebra related to the so-called $t^{3}$ model, namely $%
\mathbb{R}$, has rank $1$ (see eighth row of Table 1).}.

It is worth observing that the fourth column of Table 1, pertaining to $D=3$%
, is composed only by suitable non-compact, real forms of
\textit{all} exceptional (finite-dimensional, as understood
throughout) Lie groups (once again, with the exception of $stu$
model). Interestingly, \textit{all} exceptional Lie groups share the
property that there exists a \textit{unique} singlet in the
completely symmetric rank-$4$ tensor product of their adjoint
irrepr., namely:
\begin{equation}
\exists !~\mathbf{1}\in \left( \mathbf{Adj}\right) _{s}^{4}.
\label{res-2}
\end{equation}
On the other hand, for all infinite sequences of classical Lie
algebras (but the groups $SO\left( 8\right) $ and $SU\left( 3\right)
$), there instead exist two such singlets, \textit{i.e.}:
\begin{equation}
\exists ~\mathbf{1}_{1},\mathbf{1}_{2}\in \left( \mathbf{Adj}\right)
_{s}^{4}.  \label{2-singlets}
\end{equation}

By looking at the $G_{4}$ and $G_{3}$ given in Table 1, one can
observe that the Lie groups for which the result (\ref{res-2}) is
valid are nothing but, in suitable non-compact real forms, the
$G_{3}$'s of supergravity theories based on \textit{irreducible}
rank-$3$ Euclidean Jordan algebras (with the only exception of the
reducible, but \textit{triality symmetric}, rank-$3$ Jordan algebra
$\mathbb{R}\oplus \mathbb{R}\oplus \mathbb{R}$).

As mentioned, in the cases with $8$ supersymmetries, $G_{3}$'s and
$G_{4}$'s are related through $c$-map \cite{CFG-1}. It is also worth
remarking that, as yielded by Table 1, the unique, exceptional group
which is a $U$-duality group both in $D=4$ and in $D=3$ is $E_{7}$,
actually through all its possible non-compact, real forms, namely:
$E_{7\left(
7\right) }$ (split, \textit{i.e.} maximally non-compact, form) for maximal ($%
\mathcal{N}=8$) theory in $D=4$, $E_{7\left( -25\right) }$ for
$\mathcal{N}=2 $ ``magic''\textit{\ }octonionic model in $D=4,$ and
$E_{7\left( -5\right) }$
for $\mathcal{N}=4$ ``magic''\textit{\ (dual} to $\mathcal{N}=12$\textit{) }%
quaternionic supergravity in $D=3$.

As anticipated, exceptions to (\ref{2-singlets}) are provided by the
following classical groups:
\begin{eqnarray}
SO\left( 8\right)  &:&\exists ~\mathbf{1}_{1},\mathbf{1}_{2},\mathbf{1}%
_{3}\in \left( \mathbf{28}\right) _{s}^{4};  \label{SO(8)} \\
SU\left( 3\right)  &:&\exists !~\mathbf{1}\in \left(
\mathbf{8}\right) _{s}^{4}.  \label{SU(3)}
\end{eqnarray}

The three singlets characterizing the case of $SO\left( 8\right) $,
which appears as $D=3$ $U$-duality group of the $stu$ model through
its non-compact form $SO\left( 4,4\right) $, can actually be traced
back to the \textit{triality} of $SO\left( 8\right) $ itself
(related to the three-fold symmetry of its Dynkin diagram).

On the other hand, $SU\left( 3\right) $, in its non-compact form
$SU\left( 2,1\right) $, is the $D=3$ $U$-duality group of the
so-called ``universal hypermultiplet'' scalar sector, parameterized
by
\begin{equation}
\frac{SU\left( 2,1\right) }{SU\left( 2\right) \times U\left(
1\right) },
\end{equation}
which is both a rank-$1$ special K\"{a}hler and quaternionic
manifold of
real dimension $4$, obtained as the $c$-map \label{CFG-1} of the ``pure'' $%
\mathcal{N}=2$, $D=4$ supergravity. It is an example of Einstein
space with self-dual Weyl curvature \cite{Bagger-Witten}.

Observation (\ref{res-2}) allows us to prove a crucial identity
involving four structure constants, holding for \textit{all}
exceptional groups:
\begin{equation}
f_{\alpha \epsilon \tau }f_{\beta }{}^{\epsilon \rho }f_{\gamma
}{}^{\tau \sigma }f_{\delta \rho \sigma }=a[g_{\alpha \delta
}g_{\beta \gamma }+2g_{\alpha (\beta }g_{\gamma )\delta
}]+b[2f^{\epsilon }{}_{\alpha \gamma }f_{\epsilon \beta \delta
}-f^{\epsilon }{}_{\alpha \delta }f_{\epsilon \beta \gamma }],
\label{eqab}
\end{equation}
where $a$ and $b$ are real ($G$-dependent) constants to be
determined.

In order to prove (\ref{eqab}), we start by noticing that the
expression on its right-hand side is symmetric upon the exchanges
$\alpha \leftrightarrow \delta $ and $\beta \leftrightarrow \gamma
$, as well as upon the simultaneous exchanges $\alpha
\leftrightarrow \beta $, $\gamma \leftrightarrow \delta $.
Therefore, the indices can either be completely symmetric or with
mixed symmetry (such that the complete symmetrisation of any three
indices vanishes). The completely symmetric part is the term of the
right-hand side of (\ref{eqab}) proportional to $a$, and this is
fixed by the property (\ref{res-2}). On the other hand, the mixed
symmetry part is the term of the right-hand side of (\ref{eqab})
proportional to $b$, which is then determined using the Jacobi
identity
\begin{equation}
f_{\alpha \lbrack \beta \gamma }f^{\alpha }{}_{\delta ]\sigma }=0\,.
\label{JI}
\end{equation}
For later convenience, let us define the ($G$-dependent) constant
\begin{equation}
k\equiv \frac{g^{\vee }}{\widetilde{I}},
\end{equation}
such that \textit{e.g.} the identity (\ref{ff=deltacasimir}) can be
rewritten as
\begin{equation}
f_{\alpha \gamma \delta }f_{\beta }^{~\gamma \delta }=-kg_{\alpha
\beta }. \label{ff=kg}
\end{equation}
By suitably contracting indices, it is then straightforward to
obtain:
\begin{equation}
a=\frac{5k^{2}}{6\left( d+2\right) },\qquad b=-\frac{1}{6}k,
\end{equation}
which plugged into (\ref{eqab}) leads to
\begin{equation}
f_{\alpha \epsilon \tau }f_{\beta }{}^{\epsilon \rho }f_{\gamma
}{}^{\tau \sigma }f_{\delta \rho \sigma }=\frac{5k^{2}}{6\left(
d+2\right) }[g_{\alpha
\delta }g_{\beta \gamma }+2g_{\alpha (\beta }g_{\gamma )\delta }]-\frac{1}{6}%
k[2f^{\epsilon }{}_{\alpha \gamma }f_{\epsilon \beta \delta
}-f^{\epsilon }{}_{\alpha \delta }f_{\epsilon \beta \gamma
}].~\blacksquare   \label{ffff}
\end{equation}

The identity (\ref{ffff}) was originally determined for $E_{8}$ in
\cite {E8conventions} by using computer manipulations. The present
analysis shows that the same identity applies to \textit{all}
exceptional Lie groups. The values of the constants $k$, $a$ and $b$
appearing in the above identities, as well as the values of $g^{\vee
}$, $\tilde{I}$, $d$ and $f$, are summarised for all exceptional
groups in Table 2. As well known, $E_{8}$ is peculiar, because
$\mathbf{Adj}=\mathbf{Fund}\left( =\mathbf{248}\right) $ for this
group.
\begin{table}[tbp]
\begin{center}
\begin{tabular}{|c||c|c|c|c|c|c|c|}
\hline \rule[-1mm]{0mm}{6mm} $G$ & $g^{\vee }$ & $\tilde{I}$ & $d$ &
$f$ & $k$ & $a$ & $b$ \\ \hline\hline \rule[-1mm]{0mm}{6mm} $G_{2}$
& $4$ & $1$ & $14$ & $7$ & $4$ & ${\frac{5}{6}} $ & $-{\frac{2}{3}}$
\\ \hline
\rule[-1mm]{0mm}{6mm} $F_{4}$ & $9$ & $3$ & $52$ & $26$ & $3$ & ${\frac{5}{36%
}}$ & $-{\frac{1}{2}}$ \\ \hline
\rule[-1mm]{0mm}{6mm} $E_{6}$ & $12$ & $3$ & $78$ & $27$ & $4$ & ${\frac{1}{6%
}}$ & $-{\frac{2}{3}}$ \\ \hline
\rule[-1mm]{0mm}{6mm} $E_{7}$ & $18$ & $6$ & $133$ & $56$ & $3$ & ${\frac{1}{%
18}}$ & $-{\frac{1}{2}}$ \\ \hline
\rule[-1mm]{0mm}{6mm} $E_{8}$ & $30$ & $30$ & $248$ & $248$ & $1$ & ${\frac{1%
}{300}}$ & $-{\frac{1}{6}}$ \\ \hline
\end{tabular}
\end{center}
\caption{Table giving the dual Coxeter number $g^{\vee }$, the Dynkin index $%
\tilde{I}$ (and their ratio $k$), the dimensions $d$ and $f$, as
well as the parameters $a$ and $b$, for all exceptional Lie groups }
\label{grouptheorytable}
\end{table}

Another identity with three structure constants, exploited in Sects.
\ref {E8--->E7} and \ref{E7--->E6}, is
\begin{equation}
f_{\alpha \delta \epsilon }f_{\beta ~\rho }^{~\delta }f_{\gamma
}{}^{\epsilon \rho }=-\frac{k}{2}f_{\alpha \beta \gamma },
\label{threef}
\end{equation}
which can be proved using the Jacobi identities (\ref{JI}).

Starting from $E_{8}$, in Sects. \ref{E8--->E7} and \ref{E7--->E6},
identities (\ref{JI}), (\ref{ff=kg}), (\ref{ffff}) and
(\ref{threef}) will be used to derive many relevant identities of
$E_{7}$ and $E_{6}$ which
involve the invariant tensors made out of the $\mathbf{Fund}$ and $\mathbf{%
Adj}$ irreprs. occurring in the reduction. Our procedure amounts to
splitting the indices of $\mathbf{Fund}$ and $\mathbf{Adj}$ with
respect to the relevant maximal symmetric group embeddings, and then
to analyzing the invariant tensor structures occurring in the
branching of products of irreprs..

As discussed above, such a a progressive reduction of the
$U$-duality groups corresponds to a progressive oxidation, namely to
a progressive uplift of the space-time dimension $D$ in which the
corresponding supergravity theory is defined.

\section{\label{E8--->E7}$E_{8}\supset E_{7}\times SU(2)$}

The aim of this section and the next one is to determine all
possible $E_7$ and $E_6$ identities that result from the $E_8$
identities listed in the previous section. Most of these identities
are already known in the form we write them, in particular in the
supergravity literature they have been used to derive the
constraints satisfied by the so called ``embedding tensor'', and
thus determine all possible gaugings of maximal supergravity
theories in any dimensions (see \textit{e.g.}
\cite{Trig-2,Trig-3,lediffon}). Most identities have also been
derived in \cite{Riccioni-Hierarchy}\footnote{See in particular
Appendix A of \cite{Riccioni-Hierarchy}.}, where all the possible
gauging have been determined from $E_{11}$, following the results of
\cite{Riccioni-origin} (see also \cite{trombone}, where the so
called ``trombone'' gaugings of \cite{lediffon} were shown to result
from $E_{11}$). Still, we are now aware of the appearance in the
literature of some of the identities we list, like the
$E_7$-identities (\ref{tttt=gg+ffE7}) and (\ref{tttt=CC+tt}), and
the $E_6$-identity (\ref{PA-2-pre}). Anyway, what we want to
emphasise the most here is the straightforward $E_8$ origin of all
identities derived in this Section and in the next one.

In this Section, we consider the maximal and symmetric group embedding%
\footnote{%
Unless otherwise noted, all group embeddings considered in the
present paper are maximal and symmetric.}
\begin{equation}
E_{8}\supset E_{7}\times SU(2),  \label{emb-1}
\end{equation}
and we derive all the $E_{7}$-identities arising from the
corresponding branching of the $E_{8}$-identities (\ref{JI}),
(\ref{ff=kg}), (\ref{ffff}) and (\ref{threef}).

As mentioned above, $E_{8}$ is a peculiar exceptional Lie group, because $%
\mathbf{Adj}=\mathbf{Fund}\left( =\mathbf{248}\right) $. From the
theory of
symmetric invariant tensors of the $\mathbf{Adj}$ of Lie groups (see \textit{%
e.g.} \cite{FS-book}), it is known that the $\mathbf{248}$ of
$E_{8}$ admit
eight invariant tensors of order $2$, $8$, $12$, $14$, $18$, $20$, $24$ and $%
30$. The order-$2$ and order-$8$ invariants correspond to
\textit{primitive} invariant tensors, in terms of which the higher
ones should be expressible \cite{octic}. The quadratic one is
nothing but the Cartan-Killing metric, whereas the octic one has
been recently constructed (for $E_{8\left(
-248\right) }$ and its split form $E_{8\left( 8\right) }$, in a manifestly $%
Spin\left( 16\right) /\mathbb{Z}_{2}$-covariant form) in
\cite{octic}. To the best of our knowledge, explicit expressions of
all other higher-order invariants (also in terms of the rank-$2$ and
rank-$8$ invariants) are currently unavailable. However, this will
not affect the subsequent analysis, in which only the
$E_{8}$-invariant tensors given by the rank-$2$ (symmetric)
Cartan-Killing metric and by the rank-$3$ (completely antisymmetric)
structure constants are involved.

Under (\ref{emb-1}), the $\mathbf{248}$ of $E_{8}$ branches as
\begin{equation}
\mathbf{248}\rightarrow (\mathbf{133},\mathbf{1})+(\mathbf{56},\mathbf{2})+(%
\mathbf{1},\mathbf{3})\,,  \label{248is133plusetc}
\end{equation}
where $\mathbf{133}=\mathbf{Adj}\left( E_{7}\right) $ and $\mathbf{56}=%
\mathbf{R}\left( E_{7}\right) =\mathbf{Fund}\left( E_{7}\right) $.

We will denote the indices in $\mathbf{248}$ of $E_{8}$ with tilded
Greek indices $\widetilde{\alpha },\widetilde{\beta },...$, whereas
the indices in $\mathbf{133}$ and $\mathbf{56}$ of $E_{7}$ will be
denoted by Greek indices $\alpha ,\beta ,...$ and capital Latin
indices $M,N,...$, respectively. The
index $i=1,2,3$ and the index $a=1,2$ respectively denote the $\mathbf{3}=%
\mathbf{Adj}$ (spin $s=1$) and $\mathbf{2}=\mathbf{Fund}$ (spin $s=1/2$ ) of%
\footnote{%
A summary of our $SU(2)$ conventions is given in App. \ref{SU(2)-Convs}.} $%
SU(2)$. Within these notations, the index splitting induced by (\ref
{248is133plusetc}) reads as
\begin{equation}
\widetilde{\alpha }\rightarrow \left( \alpha ,~~Ma\ ,~~i\right) .
\end{equation}

The Cartan-Killing metric $g_{\widetilde{\alpha }\widetilde{\beta }}$ of $%
E_{8}$ branches according to
\begin{equation}
g_{\widetilde{\alpha }\widetilde{\beta }}\rightarrow \left(
g_{\alpha \beta }\,,\ \ \mathbb{C}_{MN}\epsilon _{ab}\,,\ \
g_{ij}\right) , \label{reductionmetricE8}
\end{equation}
where $\mathbb{C}_{MN}$ is the symplectic invariant metric of
$E_{7}$ (indeed, the $\mathbf{56}$ is symplectic; recall
(\ref{sympl-def})), satisfying
\begin{equation}
\mathbb{C}^{MN}\mathbb{C}_{NP}=-\delta _{P}^{M}.  \label{CC=-delta}
\end{equation}

Concerning the decomposition of the $E_{8}$ structure constants $f_{%
\widetilde{\alpha }\widetilde{\beta }\widetilde{\gamma }}$ according
to (\ref {248is133plusetc}), one should notice that in this case the
normalisation in the reduction is not free, because Eq.
(\ref{ff=kg}) relates it with the
normalisation of the Cartan-Killing metric. Thus, the normalisation used in (%
\ref{reductionmetricE8}) constrains the normalisation in the reduction of $%
f_{\widetilde{\alpha }\widetilde{\beta }\widetilde{\gamma }}$ :
\begin{equation}
f_{\widetilde{\alpha }\widetilde{\beta }\widetilde{\gamma
}}\rightarrow \left( a\ f_{\alpha \beta \gamma }\,,\ \ b\
D_{i,ab}\mathbb{C}_{MN}\,,\ \ c\ t_{\alpha \mid MN}\epsilon
_{ab}\,,\ \ d\ \epsilon _{ijk}\right) , \label{reductionstructureE8}
\end{equation}
where $a$, $b$, $c$ and $d$ are real parameters to be determined,
and where
\begin{equation}
t_{\alpha \mid MN}\equiv t_{\alpha \mid M}{}^{P}\mathbb{C}_{PN}
\label{DefGen}
\end{equation}
is symmetric in $MN$.

For clarity's sake, let us recall here the various identities,
discussed on general ground in Sect. \ref{Sect-2}, and decomposed,
in the case of $E_{8}$ under (\ref{emb-1}), in the treatment below:

\begin{enumerate}
\item  Jacobi identity:
\begin{equation}
f_{\widetilde{\alpha }[\widetilde{\beta }\widetilde{\gamma }}f^{\widetilde{%
\alpha }}{}_{\widetilde{\delta }]\widetilde{\sigma }}=0;
\label{JIE8}
\end{equation}

\item  definition of the Cartan-Killing metric:
\begin{equation}
g_{\widetilde{\alpha }\widetilde{\beta }}=-f_{\widetilde{\alpha }\widetilde{%
\gamma }\widetilde{\delta }}f_{\widetilde{\beta }}{}^{\widetilde{\gamma }%
\widetilde{\delta }};  \label{metricE8}
\end{equation}

\item  the identity with three structure constants:
\begin{equation}
f_{\widetilde{\alpha }\widetilde{\delta }\widetilde{\epsilon }}f_{\widetilde{%
\beta }}{}^{\widetilde{\delta }}{}_{\widetilde{\rho }}f_{\widetilde{\gamma }%
}{}^{\widetilde{\epsilon }\widetilde{\rho }}=-\frac{1}{2}f_{\widetilde{%
\alpha }\widetilde{\beta }\widetilde{\gamma }};  \label{threefE8}
\end{equation}

\item  the identity with four structure constants:
\begin{equation}
f_{\widetilde{\alpha }\widetilde{\epsilon }\widetilde{\tau }}f_{\widetilde{%
\beta }}{}^{\widetilde{\epsilon }\widetilde{\rho }}f_{\widetilde{\gamma }%
}{}^{\widetilde{\tau }\widetilde{\sigma }}f_{\widetilde{\delta }\widetilde{%
\rho }\widetilde{\sigma }}=\frac{1}{300}[g_{\widetilde{\alpha }\widetilde{%
\delta }}g_{\widetilde{\beta }\widetilde{\gamma }}+2g_{\widetilde{\alpha }(%
\widetilde{\beta }}g_{\widetilde{\gamma })\widetilde{\delta }}]-\frac{1}{6}[%
2f^{\widetilde{\epsilon }}{}_{\widetilde{\alpha }\widetilde{\gamma }}f_{%
\widetilde{\epsilon }\widetilde{\beta }\widetilde{\delta }}-f^{\widetilde{%
\epsilon }}{}_{\widetilde{\alpha }\widetilde{\delta
}}f_{\widetilde{\epsilon }\widetilde{\beta }\widetilde{\gamma }}].
\label{fourfE8}
\end{equation}
\end{enumerate}

We now proceed with the reduction of such $E_{8}$-identities under
the embedding (\ref{emb-1})-(\ref{248is133plusetc}).

\begin{enumerate}
\item  Let us start with the reduction of the $E_{8}$-Jacobi identity (\ref
{JIE8}). If all free indices are either in the $\mathbf{133}$ of
$E_{7}$ or in the $\mathbf{3}$ of $\left( SU(2)\right) $, one simply
gets the corresponding Jacobi identity. If three indices are in the
$\mathbf{133}$, this implies the previous case. The first
non-trivial case corresponds to having two and only two indices in
the $\mathbf{133}$, implying that the
other two indices are in the $\mathbf{56}$ of $E_{7}$; thus, one obtains (%
\ref{tt=ft}) for the generators of $E_{7}$, provided that
\begin{equation}
c=-a.
\end{equation}
Similarly, if two indices are in the $\mathbf{3}$ of $SU(2)$, one
gets
\begin{equation}
d=-b.
\end{equation}
All the terms with only one index in the $\mathbf{Adj}$ of either
$E_{7}$ or $SU(2)$ identically vanish. Finally, the case in which
all indices are in the $\mathbf{56}$ of $E_{7}$ should be
considered. Using the Fierz identity
\begin{equation}
\epsilon _{\lbrack ab}\epsilon _{c]d}=0,  \label{twoepsilonFierz}
\end{equation}
one gets
\begin{equation}
\begin{array}{l}
\epsilon _{ab}\epsilon _{cd}\left[
\begin{array}{l}
c^{2}t_{\alpha \mid MN}t_{PQ}^{\alpha }-c^{2}t_{\alpha \mid
PM}t_{NQ}^{\alpha } \\
-\frac{b^{2}}{4}\mathbb{C}_{MN}\mathbb{C}_{PQ}+\frac{b^{2}}{2}\mathbb{C}_{NP}%
\mathbb{C}_{MQ}-\frac{b^{2}}{4}\mathbb{C}_{PM}\mathbb{C}_{NQ}
\end{array}
\right]  \\
\\
+\epsilon _{bc}\epsilon _{ad}\left[
\begin{array}{l}
c^{2}t_{\alpha \mid NP}t_{MQ}^{\alpha }-c^{2}t_{\alpha \mid
PM}t_{NQ}^{\alpha } \\
+\frac{b^{2}}{4}\mathbb{C}_{NP}\mathbb{C}_{MQ}-\frac{b^{2}}{2}\mathbb{C}_{MN}%
\mathbb{C}_{PQ}+\frac{b^{2}}{4}\mathbb{C}_{PM}\mathbb{C}_{NQ}
\end{array}
\right] =0.
\end{array}
\label{PA-1}
\end{equation}
Given that the two terms in (\ref{PA-1}) are independent, this implies the $%
E_{7}$-identity
\begin{equation}
c^{2}t_{\alpha \mid M[N}t_{P]Q}^{\alpha }-\frac{b^{2}}{4}\mathbb{C}_{M[N}%
\mathbb{C}_{P]Q}+\frac{b^{2}}{4}\mathbb{C}_{NP}\mathbb{C}_{MQ}=0.
\label{tt=CCwithbandc}
\end{equation}

\item  We now perform the exceptional reduction of (\ref{metricE8}), by
recalling the branching (\ref{reductionmetricE8}). The $g_{\alpha
\beta }$ term yields the identities (\ref{ourmetric}) and
(\ref{ff=kg}) for $E_{7}$, provided that
\begin{equation}
a^{2}=\frac{1}{5}.  \label{asquaredisoneoverfive}
\end{equation}
On the other hand, by using the $SU(2)$ conventions reported in App.
\ref {SU(2)-Convs}, one can show that the $g_{ij}$ term yields
\begin{equation}
b^{2}=\frac{1}{30}.  \label{bsquaredisoneoverthirty}
\end{equation}
Finally, the $\mathbb{C}_{MN}\epsilon _{ab}$ term gives
\begin{equation}
\frac{3}{2}b^{2}+\frac{19}{4}c^{2}=1,
\end{equation}
which is identically satisfied. By plugging
(\ref{asquaredisoneoverfive}) and (\ref{bsquaredisoneoverthirty})
into (\ref{tt=CCwithbandc}), one obtains the following
$E_{7}$-identity:
\begin{equation}
t_{\alpha \mid M[N}t_{P]Q}^{\alpha }-\frac{1}{24}\mathbb{C}_{M[N}\mathbb{C}%
_{P]Q}+\frac{1}{24}\mathbb{C}_{NP}\mathbb{C}_{MQ}=0.  \label{tt=CC}
\end{equation}
It should be pointed out that all the identities which can be
obtained through the ``exceptional reduction'' approach under
consideration are \textit{invariant} under simultaneous change of
sign of the generators and of the structure constants, for both
$E_{7}$ and $SU(2)$. Therefore, the coefficients in Eq.
(\ref{reductionstructureE8}) can only be determined up to an
independent sign in front of $a$ and $c$, and in front of $b$ and
$d$. By assuming $a$ and $d$ to be positive, the performed analysis
fixes $a,b,c,d $ completely, and the consistent normalization of the
branching (\ref {reductionstructureE8}) reads
\begin{equation}
f_{\widetilde{\alpha }\widetilde{\beta }\widetilde{\gamma
}}\rightarrow
\left( \tfrac{1}{\sqrt{5}}\ f_{\alpha \beta \gamma }\,,\ \ -\tfrac{1}{\sqrt{%
30}}\ D_{i,ab}\mathbb{C}_{MN}\,,\ \ -\tfrac{1}{\sqrt{5}}\ t_{\alpha
\mid MN}\epsilon _{ab}\,,\ \ \tfrac{1}{\sqrt{30}}\ \epsilon
_{ijk}\right) . \label{reductionstructureE8fixed}
\end{equation}

\item  Next, let us consider the exceptional reduction of (\ref{threefE8}).
If the free indices are $\alpha \beta \gamma $, by using the
identity (\ref {threef}) for $E_{7}$, one gets
\begin{equation}
t_{\alpha \mid M}{}^{N}t_{\beta \mid P}{}^{M}t_{\gamma \mid N}{}^{P}=-\frac{1%
}{2}f_{\alpha \beta \gamma },  \label{ttt=-1/2fE7}
\end{equation}
which can be proved by using the symmetry in ${MN}$ of the generators of $%
E_{7}$. The identity (\ref{threefE8}) is also trivially true if the
free indices are $ijk$. The unique other non-trivial identity comes
from setting one index in the $\mathbf{133}$ and the other two
indices in the $\mathbf{56} $; in such a case, (\ref{threefE8})
implies
\begin{equation}
t_{\alpha \mid N}{}^{M}t_{\beta \mid MP}t^{\beta \mid NQ}={\frac{7}{8}}%
t_{\alpha \mid P}{}^{Q}.  \label{ttt=7/8t}
\end{equation}
All other values of the free indices yield trivial relations.

\item  Finally, we consider the reduction of (\ref{fourfE8}). If all four
indices are in the $\mathbf{133}$, using the identity (\ref{threef}) for $%
E_{7}$, one obtains the $E_{7}$-identity:
\begin{equation}
t_{\alpha \mid M}{}^{N}t_{\beta \mid N}{}^{P}t_{\gamma \mid
Q}{}^{M}t_{\delta \mid P}{}^{Q}=\tfrac{1}{72}[g_{\alpha \delta
}g_{\beta \gamma }+2g_{\alpha (\beta }g_{\gamma )\delta
}]-\tfrac{1}{6}[2f^{\epsilon }{}_{\alpha \gamma }f_{\epsilon \beta
\delta }-f^{\epsilon }{}_{\alpha \delta }f_{\epsilon \beta \gamma
}]\ .  \label{tttt=gg+ffE7}
\end{equation}
On the other hand, if all indices are in the $\mathbf{3}$ of
$SU(2)$, (\ref {threefE8}) is identically satisfied using the
properties of the Pauli
matrices. Moreover, if one sets the indices $\widetilde{\alpha }$ and $%
\widetilde{\beta }$ in the $\mathbf{133}$ and the indices
$\widetilde{\gamma
}$ and $\widetilde{\delta }$ in the $\mathbf{56}$, then the following $E_{7}$%
-identity is achieved:
\begin{equation}
f_{\alpha \epsilon }{}^{\tau }f_{\beta }{}^{\epsilon \rho }t_{\tau
\mid M}{}^{P}t_{\rho \mid PN}+t_{\alpha \mid P}{}^{R}t_{\beta \mid
RQ}t_{\gamma
\mid M}{}^{P}t_{N}^{\gamma }{}^{Q}=-\tfrac{1}{12}g_{\alpha \beta }\mathbb{C}%
_{MN}-\tfrac{37}{24}t_{\alpha \mid M}{}^{P}t_{\beta \mid PN}+\tfrac{5}{6}%
t_{\beta \mid M}{}^{P}t_{\alpha \mid PN},  \label{fftt=gC+ttE7}
\end{equation}
which can equivalently be rewritten as\footnote{%
(\ref{correctsA16}) corrects a typo in the identity (A.16) of \cite
{Riccioni-Hierarchy}.}
\begin{equation}
(t^{\gamma }t_{\alpha })_{Q}{}^{R}t_{P}^{\beta }{}^{Q}t_{\beta \mid R}{}^{M}=%
{\tfrac{1}{24}}\delta _{P}^{M}\delta _{\alpha }^{\gamma }+{\tfrac{11}{24}}%
(t^{\gamma }t_{\alpha })_{P}{}^{M}-{\tfrac{5}{12}}(t_{\alpha
}t^{\gamma })_{P}{}^{M}.  \label{correctsA16}
\end{equation}
Finally, if all indices are in the $\mathbf{56}$, by using the Fierz
identity (\ref{twoepsilonFierz}) the reduction of (\ref{fourfE8})
leads to two expressions turning out to be identical after using
(\ref{tt=CC}), and resulting into the following $E_{7}$-identity :
\begin{eqnarray}
&&t_{\alpha \mid M}{}^{R}t_{\beta \mid RN}t_{P}^{\alpha
}{}^{S}t_{SQ}^{\beta }-\tfrac{1}{6}t_{\alpha \mid MP}t_{NQ}^{\alpha
}+\tfrac{1}{12}t_{\alpha \mid
MN}t_{PQ}^{\alpha }+\tfrac{5}{6}t_{\alpha \mid MQ}t_{NP}^{\alpha }  \notag \\
&&-\tfrac{19}{144}\mathbb{C}_{MP}\mathbb{C}_{NQ}-\tfrac{47}{576}\mathbb{C}%
_{MN}\mathbb{C}_{PQ}+\tfrac{7}{144}\mathbb{C}_{MQ}\mathbb{C}_{NP}=0.
\label{tttt=CC+tt}
\end{eqnarray}
\end{enumerate}

These are all the non-trivial relations among $E_{7}$-invariant
tensors that
can be obtained by performing the reduction of the $E_{8}$-identities %
\eqref{JIE8}-\eqref{fourfE8} under the embedding (\ref{emb-1})-(\ref
{248is133plusetc}). Apart from the identities \eqref{ourmetric}, %
\eqref{gDDisdoverdlambda}, \eqref{JI}, \eqref{ff=kg}, \eqref{ffff} and %
\eqref{threef}, we also derived the $E_{7}$-identities \eqref{tt=CC}, %
\eqref{ttt=-1/2fE7}, \eqref{ttt=7/8t}, \eqref{tttt=gg+ffE7}, %
\eqref{fftt=gC+ttE7} and \eqref{tttt=CC+tt}.

\section{\label{E7--->E6}$E_{7}\supset E_{6}\times U(1)$}

In this Section, we consider the group embedding
\begin{equation}
E_{7}\supset E_{6}\times U(1),  \label{emb-2}
\end{equation}
and we derive all the $E_{6}$-identities arising from the
corresponding branching of the $E_{7}$-identities obtained in Sect.
\ref{E8--->E7}, given
by \eqref{tt=CC}, \eqref{ttt=-1/2fE7}, \eqref{ttt=7/8t}, \eqref{tttt=gg+ffE7}%
, \eqref{fftt=gC+ttE7} and \eqref{tttt=CC+tt}.

In order to perform the reduction of the $E_{7}$-identities down to
$E_{6}$,
we start and decompose the $\mathbf{Adj}=\mathbf{133}$ and $\mathbf{Fund}=%
\mathbf{56}$ irreprs. of $E_{7}$ in terms of $E_{6}\times U(1)$:
\begin{eqnarray}
\mathbf{133} &\rightarrow &\mathbf{78}_{0}+\mathbf{27}_{-2}+\mathbf{%
\overline{27}}_{+2}+\mathbf{1}_{0};  \label{decomp-2} \\
\mathbf{56} &\rightarrow &\mathbf{27}_{+1}+\mathbf{\overline{27}}_{-1}+%
\mathbf{1}_{+3}+\mathbf{\overline{1}}_{-3},  \label{decomp-3}
\end{eqnarray}
where the subscripts denote the $U(1)$-charges, and the two singlets
in (\ref {decomp-3}) are written in such a way that their opposite
$U(1)$-charges are
manifest. Concerning the $E_{6}$-irreprs., we will here denote the $\mathbf{%
Adj}\left( E_{6}\right) =\mathbf{78}$ with hatted lowercase Greek indices $%
\widehat{\alpha }$, and the $\mathbf{Fund}\left( E_{6}\right)
=\mathbf{27}$ and $\overline{\mathbf{Fund}}\left( E_{6}\right)
=\mathbf{\overline{27}}$ with, covariant (lower) respectively
contravariant (upper), hatted uppercase
Latin indices $\widehat{M}$. Thus, the decompositions (\ref{decomp-2}) and (%
\ref{decomp-3}) respectively yield the following index splittings:
\begin{eqnarray}
V_{\alpha } &\rightarrow &\left( V_{\widehat{\alpha }}\ ,\ \ V_{\widehat{M}%
}\ ,\ \ V^{\widehat{M}}\ ,\ \ V_{1}\right) ; \\
W_{M} &\rightarrow &\left( V_{\widehat{M}}\ ,\ \ V^{\widehat{M}}\ ,\
\ V_{1}\ ,\ \ V_{\overline{1}}\right) .
\end{eqnarray}

From (\ref{decomp-2}), the Cartan-Killing metric of the $\mathbf{133}$ of $%
E_{7}$ decomposes according to
\begin{equation}
g_{\alpha \beta }\rightarrow \left( g_{\widehat{\alpha }\widehat{\beta }}\ (%
\mathbf{78\ 78})\ ,\ \ \delta _{\widehat{M}}^{\widehat{N}}\ (\mathbf{27\ {%
\overline{27}}})\ ,\ \ 1\ (\mathbf{1\ 1})\right) ,
\label{reductiongE7}
\end{equation}
whereas from (\ref{decomp-3}) the symplectic invariant tensor $\mathbb{C}%
_{MN}$ of the $\mathbf{56}$ of $E_{7}$ branches as
\begin{equation}
\mathbb{C}_{MN}\rightarrow \left( \delta _{\widehat{M}}^{\widehat{N}}\ (%
\mathbf{27\ {\overline{27}}}),\ \ 1\ (\mathbf{1}\ \mathbf{\overline{1}}%
)\right) .  \label{reductionCE7}
\end{equation}
In (\ref{reductiongE7}) and (\ref{reductionCE7}), we notated in
brackets the irreprs. to which the indices ($\alpha \beta $
respectively $MN$) belong in each term. Note that all terms
occurring in the reduction must be invariant
tensors of $E_{6}\times U(1)$, and thus they trivially have vanishing $U(1)$%
-charge.

As also holding for the reductions considered in Sect.
\ref{E8--->E7}, the normalisation in the reduction of the structure
constants and of the generators cannot be arbitrarily chosen,
because it is related to the
normalisation of the Cartan-Killing metric and of the invariant tensor $%
\delta _{M}^{N}$. We will fix such normalisations further below.

The generator $t_{\alpha \mid MN}$ of $E_{7}$, which is symmetric in
$MN$, is thus decomposed according to
\begin{eqnarray}
t_{\alpha \mid MN} &\rightarrow &\left(
\begin{array}{l}
a\ t_{\widehat{\alpha }\mid \widehat{M}}{}^{\widehat{N}}\ (\mathbf{78\ 27\ {%
\overline{27}}})\ ,\ \  \\
\\
b\ d_{\widehat{P}\widehat{M}\widehat{N}}\ (\mathbf{27\ 27\ 27})\ ,~-b\ d^{%
\widehat{P}\widehat{M}\widehat{N}}\ (\mathbf{{\overline{27}}\ {\overline{27}}%
\ {\overline{27}}})\ , \\
\  \\
c\ \delta _{\widehat{M}}^{\widehat{N}}\ (\mathbf{1\ 27\
{\overline{27}}})\
,\  \\
\\
d\ \delta _{\widehat{P}}^{\widehat{M}}\ (\mathbf{27\
{\overline{27}}\ 1})\
,~d\ \delta _{\widehat{M}}^{\widehat{P}}\ (\mathbf{{\overline{27}}\ 27\ {%
\overline{1}}})\ , \\
\\
e\ (\mathbf{1\ 1\ \overline{1}})
\end{array}
\right)   \notag \\
&&  \label{reductiontE7}
\end{eqnarray}
where $d_{\widehat{M}\widehat{N}\widehat{P}}$ and $d^{\widehat{M}\widehat{N}%
\widehat{P}}$ are the rank-$3$ completely symmetric invariant
tensors of the $\mathbf{27}$ and $\overline{\mathbf{27}}$ of
$E_{6}$, namely the singlets:
\begin{equation}
d_{\widehat{M}\widehat{N}\widehat{P}}\equiv \mathbf{1}\in \left( \mathbf{27}%
\right) _{s}^{3},~~~d^{\widehat{M}\widehat{N}\widehat{P}}\equiv \mathbf{1}%
\in \left( \overline{\mathbf{27}}\right) _{s}^{3}.
\label{d-tensor-def}
\end{equation}
In (\ref{reductiontE7}), we notated in brackets the representations
to which the indices $\alpha MN$ belong for each term. The real
parameters $a,b,c,d,e$ will be determined in the following
treatment. It should be here remarked
that the terms proportional to $d_{\widehat{M}\widehat{N}\widehat{P}}$ and $%
d^{\widehat{M}\widehat{N}\widehat{P}}$ in (\ref{reductiontE7}) have
opposite coefficients, for consistency with the condition
\begin{equation}
d_{\widehat{M}\widehat{N}\widehat{P}}d^{\widehat{M}\widehat{N}\widehat{Q}%
}=\delta _{\widehat{P}}^{\widehat{Q}},  \label{ass-1}
\end{equation}
that we assume\footnote{%
This simplifying assumption changes the normalization of the $d$-tensors $d_{%
\widehat{M}\widehat{N}\widehat{P}}$ and $d^{\widehat{M}\widehat{N}\widehat{P}%
}$ with respect the one usually adopted in supergravity. For
example, let us consider the embedding (particular non-compact real
form of (\ref{emb-2}))
\begin{equation*}
E_{7(-25)}\supset E_{6(-26)}\times SO\left( 1,1\right) ,
\end{equation*}
pertaining to $\mathcal{N}=2$, $D=4$ octonionic ``magic''
supergravity
(based on $J_{3}^{\mathbb{O}}$) branched with respect to its $D=5$ $U$%
-duality group $E_{6\left( -26\right) }$. In this case, it holds
that \cite {CVP} (see also \cite{Raju-1} for recent treatment):
\begin{equation*}
d_{\widehat{M}\widehat{N}\widehat{P}}d^{\widehat{M}\widehat{N}\widehat{Q}%
}=10\delta _{\widehat{P}}^{\widehat{Q}}.
\end{equation*}
Thus, assumption (\ref{ass-1}) amounts to changing such a
normalization, and
setting to $1$ the coefficient of proportionality between $d_{\widehat{M}%
\widehat{N}\widehat{P}}d^{\widehat{M}\widehat{N}\widehat{Q}}$ and $\delta _{%
\widehat{P}}^{\widehat{Q}}$.}. Again, as was the case for %
\eqref{reductiongE7} and \eqref{reductionCE7}, the $U(1)$-charge of
each term in the decomposition \eqref{reductiontE7} vanishes.

Furthermore, the structure constants of $E_{7}$ decompose according
to
\begin{equation}
f_{\alpha \beta \gamma }\rightarrow \left( f\ f_{\widehat{\alpha }\widehat{%
\beta }\widehat{\gamma }}\ (\mathbf{78\ 78\ 78})\ ,\ \ g\
t_{\widehat{\alpha }\mid \widehat{M}}{}^{\widehat{N}}\ (\mathbf{78\
27\ {\overline{27}}})\ ,\ \
h\ \delta _{\widehat{M}}^{\widehat{N}}\ (\mathbf{27\ {\overline{27}}\ 1}%
)\right) ,  \label{reductionfE7fgh}
\end{equation}
with the real parameters $f,g,h$ to be determined.\medskip

\begin{enumerate}
\item  We start by considering the reduction of the identities that do not
involve the structure constants $f_{\alpha \beta \gamma }$ of
$E_{7}$. The
identities \eqref{ourmetric} and \eqref{gDDisdoverdlambda} specified for $%
E_{7}$ give the same relations for $E_{6}$, provided that
\begin{equation}
a^{2}=\frac{1}{2},
\end{equation}
and that the other parameters in \eqref{reductiontE7} satisfy the
system
\begin{equation}
\left\{
\begin{array}{l}
1=b^{2}+2d^{2}; \\
1=54c^{2}+2e^{2}; \\
\frac{19}{8}=27d^{2}+e^{2},
\end{array}
\right.
\end{equation}
which leaves one parameter undetermined.

\item  The reduction of the identity \eqref{tt=CC} constrains the squares of
all the parameters in the decomposition \eqref{reductiontE7} to be
\begin{equation}
b^{2}=\frac{5}{6},\quad c^{2}=\frac{1}{72},\quad
d^{2}=\frac{1}{12},\quad e^{2}=\frac{1}{8},
\end{equation}
and it also yields the further constraint
\begin{equation}
ce=\frac{1}{24}.
\end{equation}
On the other hand, the only non-trivial $E_{6}$-identity that it
produces reads
\begin{equation}
t_{\widehat{\alpha }\mid \widehat{M}}{}^{\widehat{N}}t_{~\widehat{P}}^{%
\widehat{\alpha }}{}^{\widehat{Q}}=\frac{1}{6}\delta _{\widehat{M}}^{%
\widehat{Q}}\delta _{\widehat{P}}^{\widehat{N}}+\frac{1}{18}\delta _{%
\widehat{M}}^{\widehat{N}}\delta _{\widehat{P}}^{\widehat{Q}}-\frac{5}{3}d_{%
\widehat{M}\widehat{P}\widehat{R}}d^{\widehat{N}\widehat{Q}\widehat{R}}.
\label{ttdeltadeltaddE6}
\end{equation}

\item  We then consider the reduction of the identity \eqref{ttt=7/8t},
which gives rise to the two following $E_{6}$-identities:
\begin{eqnarray}
&&t_{\widehat{M}}^{\widehat{\alpha }}{}^{\widehat{N}}d^{\widehat{M}\widehat{P%
}\widehat{Q}}d_{\widehat{N}\widehat{S}\widehat{Q}}=-\frac{1}{2}t_{\widehat{S}%
}^{\widehat{\alpha }}{}^{\widehat{P}};  \notag \\
&&t_{\widehat{M}}^{\widehat{\alpha }}{}^{\widehat{N}}t_{\widehat{\alpha }%
\mid \widehat{P}}{}^{\widehat{Q}}d_{\widehat{N}\widehat{Q}\widehat{R}}=-%
\frac{13}{9}d_{\widehat{M}\widehat{P}\widehat{R}}.  \label{twos}
\end{eqnarray}
By using the values of the parameters obtained above, it can be
checked that all other combinations of indices yield trivial
relations.

\item  The remaining $E_{7}$-identities that do not involve the structure
constants $f_{\alpha \beta \gamma }$ are given by \eqref{correctsA16} and %
\eqref{tttt=CC+tt}. The identity \eqref{correctsA16} has two free
indices in the $\mathbf{133}$ and two other ones in the
$\mathbf{56}$ of $E_{7}$. Its reduction produces only two
non-trivial $E_{6}$-identities, namely:
\begin{eqnarray}
&&\frac{1}{4}( t_{\widehat{\alpha }}t_{\widehat{\beta }}) _{%
\widehat{P}}{}^{\widehat{Q}}t_{\widehat{\gamma }\mid \widehat{M}}{}^{%
\widehat{P}}t_{~\widehat{Q}}^{\widehat{\gamma }}{}^{\widehat{N}}+\frac{5}{12}%
(t_{\widehat{\beta }}t_{\widehat{\alpha }})_{\widehat{P}}{}^{\widehat{Q}}d_{%
\widehat{M}\widehat{Q}\widehat{R}}d^{\widehat{N}\widehat{P}\widehat{R}}-%
\frac{2}{9}( t_{\widehat{\alpha }}t_{\widehat{\beta }}) _{%
\widehat{M}}{}^{\widehat{N}}  \notag \\
&&+\frac{5}{24}(t_{\widehat{\beta }}t_{\widehat{\alpha }})_{\widehat{M}}{}^{%
\widehat{N}}-\frac{1}{24}g_{\widehat{\alpha }\widehat{\beta }}\delta _{%
\widehat{M}}^{\widehat{N}}=0,
\end{eqnarray}
where the free indices are $\widehat{\alpha }$, $\widehat{\beta }$, $%
\widehat{M}$ and $\widehat{N}$ (thus, two in the $\mathbf{78}$, one in the $%
\mathbf{27}$, and the other one in $\overline{\mathbf{27}}$), and
\begin{equation}
\frac{5}{2}d_{\widehat{M}\widehat{R}\widehat{S}}d^{\widehat{N}\widehat{R}%
\widehat{T}}d_{\widehat{P}\widehat{T}\widehat{U}}d^{\widehat{Q}\widehat{S}%
\widehat{U}}+d_{\widehat{M}\widehat{P}\widehat{R}}d^{\widehat{N}\widehat{Q}%
\widehat{R}}-\frac{1}{4}\delta _{\widehat{M}}^{(\widehat{N}}\delta _{%
\widehat{P}}^{\widehat{Q})}=0,
\end{equation}
with two indices in the $\mathbf{27}$ and two other ones in the $\overline{%
\mathbf{27}}$. By using \eqref{ttdeltadeltaddE6}, one can show that
all
other identities resulting from the reduction of \eqref{correctsA16} and %
\eqref{tttt=CC+tt} are equivalent to the the ones given above.

\item  We then proceed to considering the reduction of the identities
containing the structure constants $f_{\alpha \beta \gamma }$ of
$E_{7}$. In
order to derive the coefficients $f$, $g$ and $h$ in the decomposition %
\eqref{reductionfE7fgh}, it is sufficient to consider the reduction of the $%
E_{7}$-identity \eqref{tt=ft}. This yields: $i$) again %
\eqref{reductionfE7fgh} for $E_{6}$; $ii$) the identity
\begin{equation}
t_{\widehat{\alpha }\mid (\widehat{M}}{}^{\widehat{Q}}d_{\widehat{N}\widehat{%
P})\widehat{Q}}=0,  \label{invconditionofd}
\end{equation}
which is the condition of invariance of $d_{\widehat{M}\widehat{N}\widehat{P}%
}$ itself; $iii$) the identity \eqref{ttdeltadeltaddE6}. Thus, the
parameters $f$, $g$ and $h$ must satisfy
\begin{equation}
f=a,\qquad g=-a,\qquad hc=\frac{1}{36}.
\end{equation}

\item  The reduction of the Jacobi identity of $E_{7}$ and the reduction of
the $E_{7}$-identity \eqref{ff=kg} produce no further new $E_{6}$%
-identities, while the reduction of the $E_{7}$-identity
\eqref{threef} yields the $E_{6}$-identity
\begin{equation}
t_{\widehat{\alpha }\mid
\widehat{M}}{}^{\widehat{N}}t_{\widehat{\beta }\mid
\widehat{P}}{}^{\widehat{M}}t_{\widehat{\gamma }\mid \widehat{N}}{}^{%
\widehat{P}}=-\frac{1}{2}f_{\widehat{\alpha }\widehat{\beta
}\widehat{\gamma }},
\end{equation}
together with \eqref{threef} for $E_{6}$. Also, it can be checked
that the reduction of \eqref{ttt=-1/2fE7} gives no additional
$E_{6}$-identities.

\item  Finally, we consider the reduction of the $E_{7}$-identities %
\eqref{eqab}, \eqref{tttt=gg+ffE7} and \eqref{fftt=gC+ttE7}. These
all give rise to the following two $E_{6}$-identities :
\begin{eqnarray}
&&t_{\widehat{\alpha }\mid \widehat{M}}{}^{\widehat{N}}t_{\widehat{\beta }%
\mid \widehat{P}}{}^{\widehat{M}}t_{\widehat{\gamma }\mid \widehat{N}}{}^{%
\widehat{Q}}t_{\widehat{\delta }\mid \widehat{Q}}{}^{\widehat{P}}-\frac{1}{12%
}g_{(\widehat{\alpha }\widehat{\beta }}g_{\widehat{\gamma }\widehat{\delta })}+%
\frac{1}{6}(2f^{\widehat{\epsilon }}{}_{\widehat{\alpha }\widehat{\gamma }%
}f_{\ \widehat{\epsilon }\widehat{\beta }\widehat{\delta }}-f^{\widehat{%
\epsilon }}{}_{\widehat{\alpha }\widehat{\delta }}f_{\widehat{\epsilon }%
\widehat{\beta }\widehat{\gamma }})=0;  \label{PA-2-pre} \\
&&  \notag \\
&&f^{\widehat{\epsilon }}{}_{\widehat{\alpha }\widehat{\tau }}f_{\widehat{%
\epsilon }\widehat{\beta }\widehat{\sigma }}(t^{\widehat{\tau }}t^{\widehat{%
\sigma }})_{\widehat{M}}{}^{\widehat{N}}-(t_{\widehat{\alpha }}t_{%
\widehat{\beta }}) _{\widehat{P}}{}^{\widehat{Q}}t_{\widehat{\gamma }%
\mid \widehat{M}}{}^{\widehat{P}}t_{\widehat{Q}}^{\widehat{\gamma }}{}^{%
\widehat{N}}+\frac{17}{9}(t_{\widehat{\alpha }}t_{\widehat{\beta }%
}) _{\widehat{M}}{}^{\widehat{N}}  \notag \\
&&-(t_{\widehat{\beta }}t_{\widehat{\gamma }}) _{\widehat{M}}{}^{%
\widehat{N}}+\frac{2}{9}g_{\widehat{\alpha }\widehat{\beta }}\delta _{%
\widehat{M}}^{\widehat{N}}=0.  \label{PA-2}
\end{eqnarray}
By using \eqref{ttdeltadeltaddE6}, the identity (\ref{PA-2}) can be
equivalently rewritten as
\begin{equation}
(t_{\widehat{\alpha }}t_{\widehat{\beta }}) _{\widehat{P}}{}^{%
\widehat{Q}}d_{\widehat{M}\widehat{Q}\widehat{R}}d^{\widehat{N}\widehat{P}%
\widehat{R}}+\frac{1}{5}(t_{\widehat{\alpha }}t_{\widehat{\beta }%
}) _{\widehat{M}}{}^{\widehat{N}}-\frac{3}{10}(t_{\widehat{\beta
}}t_{\widehat{\alpha }}) _{\widehat{M}}{}^{\widehat{N}}-\frac{1}{30}g_{%
\widehat{\alpha }\widehat{\beta }}\delta
_{\widehat{M}}^{\widehat{N}}=0.
\end{equation}
\end{enumerate}

To summarise, all the $E_{6}$-identities that we have obtained
result from the $E_{7}$-identities obtained in Sect. \ref{E8--->E7},
by means of
decompositions (\ref{reductiongE7}), (\ref{reductionCE7}) and of (recall (%
\ref{reductiontE7}) and (\ref{reductionfE7fgh})):
\begin{eqnarray}
&&t_{\alpha \mid MN}\rightarrow \left(
\begin{array}{l}
\frac{1}{\sqrt{2}}t_{\widehat{\alpha }\mid \widehat{M}}{}^{\widehat{N}}\ (%
\mathbf{78\ 27\ {\overline{27}}})\ ,\ \  \\
\\
\sqrt{\frac{5}{6}}d_{\widehat{P}\widehat{M}\widehat{N}}\ (\mathbf{27\ 27\ 27}%
)\ ,~-\sqrt{\frac{5}{6}}d^{\widehat{P}\widehat{M}\widehat{N}}\ (\mathbf{{%
\overline{27}}\ {\overline{27}}\ {\overline{27}}})\ ,\  \\
\\
\frac{1}{6\sqrt{2}}\delta _{\widehat{M}}^{\widehat{N}}\ (\mathbf{1\ 27\ {%
\overline{27}}})\ ,\  \\
\\
\frac{1}{2\sqrt{3}}\delta _{\widehat{P}}^{\widehat{M}}\ (\mathbf{27\ {%
\overline{27}}\ 1})\ ,~\frac{1}{2\sqrt{3}}\delta _{\widehat{M}}^{\widehat{P}%
}\ (\mathbf{{\overline{27}}\ 27\ {\overline{1}}})\ ,\  \\
\\
\frac{1}{2\sqrt{2}}\ (\mathbf{1\ 1\ \overline{1}})
\end{array}
\right) ;  \notag \\
&& \\
&&f_{\alpha \beta \gamma }\rightarrow \left(
\begin{array}{l}
\frac{1}{\sqrt{2}}f_{\widehat{\alpha }\widehat{\beta }\widehat{\gamma }}\ (%
\mathbf{78\ 78\ 78})\ , \\
\\
-\frac{1}{\sqrt{2}}t_{\widehat{\alpha }\mid \widehat{M}}{}^{\widehat{N}}\ (%
\mathbf{78\ 27\ {\overline{27}}})\ ,~ \\
\\
\frac{1}{3\sqrt{2}}\delta _{\widehat{M}}^{\widehat{N}}\ (\mathbf{27\ {%
\overline{27}}\ 1})
\end{array}
\right) .  \notag \\
&&
\end{eqnarray}

In App. \ref{SO(10)} we proceed further, and consider some examples
of
reductions of $E_{6}$-identities with respect to its maximal subgroup $%
SO\left( 10\right) \times U\left( 1\right) $.

\section{\label{K-Tensor}The $K$-Tensor and Its \textit{``Master''} Identity}

As application of the formalism developed in previous Sects., in
order to elucidate the connections to $D=4$ Maxwell-Einstein
supergravity theories based on symmetric scalar manifolds and
related to \textit{irreducible}
Euclidean Jordan algebras (the unique exception being the \textit{%
triality-symmetric} $\mathcal{N}=2$ $stu$ model), we now derive a
fundamental (dubbed \textit{``master''}) identity, involving the unique rank-%
$4$ completely symmetric invariant tensor (named $K$-tensor) of the $0$%
-brane (black hole) charge irrepr. $\mathbf{R}$ of $G_{4}$ (see also
the treatment of \cite {Og}). Besides the importance of the
$K$-tensor for the theory of extremal black hole attractors \cite
{AM} in Maxwell-Einstein supergravities, this (hitherto unknown)
identity has potential application in the issue of the
classification of the orbits of $\mathbf{R}\left( G_{4}\right) $ in
presence of Dirac-Zwanziger-Schwinger charge quantization conditions
(especially for $\mathcal{N}=8$, $D=4$ supergravity, see \textit{\
e.g.} \cite {Krut-1,Duff-1,Duff-2,N=8-UV-finiteness}, and Refs.
therein), as well as in the study of multi-center black hole
solutions \cite {FMOSY-1,irred-1}.

\textit{At least} in $D=4$ supergravities with \textit{symmetric}
scalar manifolds $\frac{G_{4}}{H_{4}}$, the $U$-duality groups
$G_{4}$'s share the
property that the generators $t_{\alpha \mid }{}_{MN}$ (\ref{DefGen}) in $%
\mathbf{R}$ are $G_{4}$-singlets:
\begin{equation}
\exists !~t_{\alpha \mid }{}_{MN}\equiv \mathbf{1}\in
\mathbf{Adj}\times _{s}\left( \mathbf{R}\times _{s}\mathbf{R}\right)
.  \label{ASS}
\end{equation}
This can be proven explicitly by using Baker-Campbell-Hausdorff
formula (see \textit{e.g.} \cite{Reinsch-1} and Refs. therein) and
acting on the generators (\ref{DefGen}) with a generic group element
$S\equiv e^{\xi ^{\beta }t_{\beta }}\in G_{4}$. In fact, $G_{4}$
induces a transformation of the symplectic indices which is
equivalent to an inverse transformation on the adjoint index:
\begin{eqnarray}
t_{\alpha }\rightarrow S\,t_{\alpha }\,S^{-1} &=&t_{\alpha }+\xi
^{\beta _{1}}\left[ t_{\beta _{1}},t_{\alpha }\right]
+\frac{1}{2}\xi ^{\beta
_{1}}\xi ^{\beta _{2}}\left[ t_{\beta _{1}}\left[ t_{\beta _{2}},t_{\alpha }%
\right] \right] +\dots =  \notag \\
&=&t_{\alpha }-\xi ^{\beta _{1}}\left( \mathcal{T}_{\beta
_{1}}\right) _{\alpha }{}^{\sigma }t_{\sigma }+\frac{1}{2}\xi
^{\beta _{1}}\xi ^{\beta _{2}}\left( \mathcal{T}_{\beta
_{1}}\mathcal{T}_{\beta _{2}}\right) _{\alpha
}{}^{\sigma }t_{\sigma }+\dots =  \notag \\
&=&\left( e^{-\xi ^{\beta }\mathcal{T}_{\beta }}\right) _{\alpha
}{}^{\sigma }\,t_{\sigma }\equiv \left( \mathcal{S}^{-1}\right)
_{\alpha }{}^{\sigma }\,t_{\sigma }\,,  \label{G-act}
\end{eqnarray}
where the adjoint representation of generators (made out of the
structure constants $f_{\alpha \beta }{}^{\gamma }$ of the Lie
algebra $\frak{g}_{4}$ of $G_{4}$)
\begin{equation}
\left( \mathcal{T}_{\alpha }\right) _{\sigma }{}^{\delta }\equiv
-f_{\alpha \sigma }{}^{\delta }\,,
\end{equation}
was used. Consequently, $\mathcal{S}\equiv e^{\xi ^{\beta }\mathcal{T}%
_{\beta }}$, defined by the last line of (\ref{G-act}), is the
adjoint representation of $S$ itself.

A key property of the symplectic representation $t^{\alpha }{}_{MN}$
of the generators is the even symmetry of its symplectic indices:
$t^{\alpha }{}_{MN}=t^{\alpha }{}_{\left( MN\right) }$. Besides
(\ref{ASS}) itself, this is also implied by the fact that $t^{\alpha
}{}_{\left[ MN\right] }$ is the symplectic variation of
$\mathbb{C}_{MN}$, which trivially vanishes, due
to the fact that the symplectic metric $\mathbb{C}_{MN}$ itself is a $G_{4}$%
-singlet (recall (\ref{sympl-def})).

By exploiting such a symmetry, it is possible to construct a
rank-$4$
completely symmetric $G_{4}$-invariant tensor of $\mathbf{R}$, dubbed $K$%
\textit{-tensor}:
\begin{equation}
\exists !\mathbb{K}_{MNPQ}\equiv \mathbf{1}\in \left(
\mathbf{R}\right) _{s}^{4},
\end{equation}
which is not a \textit{primitive} invariant, since it is in general
defined as follows:
\begin{eqnarray}
\mathbb{K}_{MNPQ}\propto t_{(MN}^{\alpha }t_{\alpha |PQ)} &=&\frac{1}{3}%
\left( t^{\alpha }{}_{MN}t_{\alpha |PQ}+t^{\alpha }{}_{MP}t_{\alpha
|QN}+t^{\alpha }{}_{MQ}t_{\alpha |PN}\right)\,.  \label{K0}
\end{eqnarray}
Such an invariant structure exists \textit{at least} in $D=4$
supergravities with \textit{symmetric} scalar manifolds
$\frac{G_{4}}{H_{4}}$. Furthermore,
for all these theories but the $\mathcal{N}=2$ $\mathbb{CP}^{n}$ \textit{%
minimal coupling} sequence \cite{Luciani} and the $\mathcal{N}=3$
theory \cite
{N=3}, the $K$%
-tensor is \textit{irreducible} in $\mathbf{R}$, namely it cannot be
expressed in terms of lower-rank tensors with indices only in
$\mathbf{R}$.

For instance, for $G_{4}=E_{7}$ (corresponding to $\mathcal{N}=8$
maximal and to $\mathcal{N}=2$ ``magic'' octonionic $D=4$
supergravity, for $E_{7(7)}
$ respectively $E_{7\left( -25\right) }$), it holds $\mathbf{R}=\mathbf{Fund}%
=\mathbf{56}$, and
\begin{equation}
\left( \mathbf{56}\right) _{s}^{4}=\overset{\mathbb{K}_{MNPQ}}{\mathbf{1}}+%
\mathbf{1463}+\mathbf{1539}+\mathbf{7371}+\mathbf{150822}+\mathbf{293930}.
\label{4-s}
\end{equation}
In this case, the $K$-tensor defined in (\ref{K0}) can be
characterized in a more useful way using the $E_{7}$-identity
(\ref{tt=CC}), which can be recast in the following form:
\begin{equation}
t^{\alpha }{}_{M\left[ N\right| }t_{\alpha |P]Q}=\frac{1}{24}\left[ \mathbb{C%
}_{M(P}\mathbb{C}_{Q)N}-\mathbb{C}_{M(N}\mathbb{C}_{Q)P}\right] \,.
\end{equation}
After some algebra, the following identity is achieved:
\begin{equation}
t_{\left( MN\right| }^{\alpha }t_{\alpha |PQ)}=t_{MN}^{\alpha
}t_{\alpha |PQ}-\frac{1}{12}\,\mathbb{C}_{M(P}\mathbb{C}_{Q)N}\,.
\label{Kalpha}
\end{equation}
By using (\ref{gDDisdoverdlambda}) and (\ref{CC=-delta}), one can
check (\ref {Kalpha}) to be skew-traceless. Thus, for $G_{4}=E_{7}$
the following fundamental relation is obtained:
\begin{equation}
\mathbb{K}_{MNPQ}=\xi \,t_{(MN}^{\alpha }t_{\alpha |PQ)}=\xi \left[
t_{MN}^{\alpha }t_{\alpha |PQ}-\frac{1}{12}\,\mathbb{C}_{M(P}\mathbb{C}_{Q)N}%
\right] \,,  \label{csi}
\end{equation}
where the real proportionality constant $\xi $ has been introduced. \textit{%
At least} in all $D=4$ supergravities with \textit{symmetric} scalar
manifolds in which the $K$-tensor is \textit{irreducible} in
$\mathbf{R}$ (see comment below (\ref{K0})), the result (\ref{csi})
can be generalized as
\begin{equation}
\mathbb{K}_{MNPQ}=\xi \left( t_{MN}^{\alpha }t_{\alpha |PQ}-\tau \,\mathbb{C}%
_{M(P}\mathbb{C}_{Q)N}\right) \,,  \label{covquan}
\end{equation}
where $\tau $ is a real constant, in general depending on $d$ and
$f$, determined by imposing the skew-tracelessness condition on
$\mathbb{K}_{MNPQ} $ (recall identities (\ref{gDDisdoverdlambda})
and (\ref{CC=-delta})):
\begin{equation}
\mathbb{C}^{NP}\mathbb{K}_{MNPQ}=0\Leftrightarrow \tau
=\frac{2d}{f(f+1)}\,. \label{tau}
\end{equation}
Thus, the following expression for the $K$-tensor is obtained:
\begin{equation}
\mathbb{K}_{MNPQ}=\xi \left[ t_{MN}^{\alpha }t_{\alpha |PQ}-\frac{2d}{f(f+1)}%
\mathbb{C}_{M(P}\mathbb{C}_{Q)N}\right] \,,  \label{K}
\end{equation}
where the real proportionality constant $\xi $ depends on the chosen
normalization of the generators $t^{\alpha }$'s, getting fixed by
the explicit computation.

Remarkably, the $K$-tensor is related to the well known invariant
homogeneous polynomial $I_{4}$ of $\mathbf{R}\left( G_{4}\right) $,
used to classify extremal black hole solutions in Maxwell-Einstein
$D=4$ supergravities with \textit{symmetric} scalar manifolds (see
\textit{e.g.} \cite{ADFT-review} for a review and a list of Refs.).
Indeed, $I_{4}$ is defined as the contraction of the $K$-tensor with
four copies of the charge vector of $\mathbf{R}$ ($\Lambda
=0,1,...,f/2-1$)
\begin{equation}
Q^{M}=(p^{\Lambda },q_{\Lambda })^{t},  \label{Q}
\end{equation}
namely:
\begin{equation}
I_{4}\equiv \mathbb{K}_{MNPQ}Q^{M}Q^{N}Q^{P}Q^{Q}=\xi t_{MN}^{\alpha
}t_{\alpha |PQ}Q^{M}Q^{N}Q^{P}Q^{Q}\,,  \label{I4}
\end{equation}
resulting in a homogeneous polynomial of degree four in the black
hole charges $Q$.

It is here worth remarking that in $D=5$ the role of the $K$-tensor
is played by the so-called $d$-tensors defined in
(\ref{d-tensor-def}), used to construct the $G_{5}$-invariant cubic
homogeneous polynomials $I_{3,e}$ (electric) and $I_{3,m}$
(magnetic) (see \textit{e.g.} \cite
{Ferrara-Maldacena,FG1,FG2,CFMZ1-d=5}). A key difference is that the $d$%
-tensor is \textit{primitive}, namely it cannot be expressed in
terms of other independent tensor structures in \textit{any}
irreprs. of $G_{5}$, while the $K$-tensor is \textit{not}
\textit{primitive} (from its very definition (\ref{K0})). Moreover,
\textit{at least} in \textit{symmetric} geometries, the $d$-tensor
satisfies the fundamental identity
\begin{equation}
d_{\widehat{M}(\widehat{N}\widehat{P}}d_{\widehat{Q}\widehat{R})\widehat{S}%
}d^{\widehat{M}\widehat{S}\widehat{T}}=\frac{2}{15}\delta _{(\widehat{N}}^{%
\widehat{T}}d_{\widehat{P}\widehat{Q}\widehat{R})}.  \label{ddd=d}
\end{equation}
This identity can be derived from the identities obtained in Sect.
\ref
{E7--->E6}, by contracting (\ref{ttdeltadeltaddE6}) with $d_{\widehat{S}%
\widehat{T}\widehat{Q}}$, and then by symmetrising with respect to
all the lower indices and using (\ref{invconditionofd}). Notice the
different normalization of (\ref{ddd=d}) \textit{e.g.} with respect
to Refs. \cite {GST,CVP} (see Footnote 11).

In analogy with the $D=5$ case, the issue of deriving an identity
analogue to (\ref{ddd=d}) involving the $K$-tensor naturally arises
out. By
exploiting the definition (\ref{K}) of the $K$-tensor and the decomposition (%
\ref{ttDec}), the following result can be achieved (recall
(\ref{tau})):
\begin{eqnarray}
&&\mathbb{K}_{MNPQ}\mathbb{K}_{RSTU}\mathbb{C}^{QR}=\mathbb{K}_{\left(
MNP\right) Q}\mathbb{K}_{R\left( STU\right) }\mathbb{C}^{QR}  \notag \\
&=&-\xi ^{2}\frac{1}{f}t_{\alpha ,\mid
(MN}\mathbb{C}_{P)(S}t_{TU)}^{\alpha }
\notag \\
&&+\xi ^{2}\frac{1}{2}f_{\alpha \beta \gamma }t_{~(MN}^{\alpha
}t_{~P)(S}^{\beta }t_{~TU)}^{\gamma }-\xi ^{2}t_{~(MN}^{\alpha
}S_{\alpha
\beta \mid P)(S}t_{~TU)}^{\beta }  \label{Master-d=4-pre} \\
&=&\xi \frac{1}{f}\mathbb{K}_{\left( MN\right|
(ST}\mathbb{C}_{U)\left| P\right) }+\xi ^{2}\frac{\tau
}{f}\mathbb{C}_{\left( M\right| \left( S\right| }\mathbb{C}_{\left|
N\right| \left| T\right| }\mathbb{C}_{\left|
P\right) \left| U\right) }  \notag \\
&&+\xi ^{2}\frac{1}{2}f_{\alpha \beta \gamma }t_{~(MN}^{\alpha
}t_{~P)(S}^{\beta }t_{~TU)}^{\gamma }-\xi ^{2}t_{~(MN}^{\alpha
}S_{\alpha \beta \mid P)(S}t_{~TU)}^{\beta },  \label{Master-d=4}
\end{eqnarray}
where $\mathbb{C}_{\left( M\right| \left( S\right|
}\mathbb{C}_{\left| N\right| \left| T\right| }\mathbb{C}_{\left|
P\right) \left| U\right) }$
means symmetrization for the triplets of indices $\left( M,N,P\right) $ and $%
\left( S,T,U\right) $. In (\ref{Master-d=4-pre}) and
(\ref{Master-d=4}) we introduced the fundamental invariant tensor
$S_{(\alpha \beta )}^{[MN]}$
(see App. \ref{Decomps}), that does not arise from the reduction of the $%
E_{8}$-identities considered in the present paper.

By some algebra, Eq. (\ref{covquan}) yields (recall (\ref{tau})):
\begin{equation}
\mathbb{K}_{MNA_{1}A_{2}}\mathbb{K}_{PQA_{3}A_{4}}\mathbb{C}^{A_{1}A_{3}}%
\mathbb{C}^{A_{2}A_{4}}=\,\xi \left[ \left( 2\tau -1\right) \mathbb{K}%
_{MNPQ}+\,\xi \,\tau \left( \tau -1\right) \mathbb{C}_{M(P}\mathbb{C}_{Q)N}%
\right] \,.  \label{KKCC}
\end{equation}
The identity (\ref{KKCC}) implies that arbitrary powers of the
$K$-tensor,
each having a couple of indices contracted, are always linear in the $K$%
-tensor and in $\mathbb{C}_{M(P}\mathbb{C}_{Q)N}$. By further
contracting with the charges $Q^{M}Q^{N}Q^{P}Q^{Q}$ and recalling
definition (\ref{I4}), one obtains
\begin{equation}
\mathbb{K}_{MNA_{1}A_{2}}\mathbb{K}_{PQA_{3}A_{4}}\mathbb{C}^{A_{1}A_{3}}%
\mathbb{C}^{A_{2}A_{4}}Q^{M}Q^{N}Q^{P}Q^{Q}=\xi \left( 2\tau
-1\right) I_{4}. \label{KKCCQQQQ-1}
\end{equation}
On the other hand, by suitably changing the order of the indices of the $K$%
-tensor and recalling Eq. (\ref{ff=deltacasimir}), one can compute
\begin{equation}
\mathbb{K}_{MA_{1}NA_{2}}\mathbb{K}_{PA_{3}QA_{4}}\mathbb{C}^{A_{1}A_{3}}%
\mathbb{C}^{A_{2}A_{4}}Q^{M}Q^{N}Q^{P}Q^{Q}=-\xi \left( \frac{g^{\vee }}{4%
\widetilde{I}}+\tau \right) I_{4}.  \label{KKCCQQQQ-2}
\end{equation}
From the complete symmetry of the $K$-tensor, the fact that the
left-hand sides of (\ref{KKCCQQQQ-1}) and (\ref{KKCCQQQQ-2}) are
equal implies the following relation:
\begin{equation}
\frac{g^{\vee }}{\widetilde{I}}=4\left( 1-3\tau \right) ,
\label{relat-1}
\end{equation}
which, through (\ref{tau}), relates the dual Coxeter number $g^{\vee
}$, the
Dynkin index of $\mathbf{R}$, and the dimensions of $\mathbf{R}$ and $%
\mathbf{Adj}$. The result (\ref{relat-1}) holds \textit{at least} for all $%
G_{4}$'s of supergravity theories reported in Table 1. For these
groups, Eqs. (\ref{ff=deltacasimir}) and (\ref{relat-1}) imply that
the general result
\begin{equation}
\frac{C_{\mathbf{Adj}}}{C_{\mathbf{R}}}=\frac{f}{d}\frac{g^{\vee }}{%
\widetilde{I}}
\end{equation}
can be further elaborated as
\begin{equation}
\frac{C_{\mathbf{Adj}}}{C_{\mathbf{R}}}=4\frac{f}{d}\left( 1-3\tau
\right) . \label{relat-2}
\end{equation}
\medskip

The \textit{``master''} identity (\ref{Master-d=4}) has potential
application in the study of \textit{independent} tensor structures in the $0$%
-brane (black hole) charge irrepr. $\mathbf{R}$ of $G_{4}$'s of \textit{%
symmetric} $D=4$ supergravities. Due to recent advances in the
investigation
of UV finiteness properties \cite{N=8-UV-finiteness}, the case of the $%
\mathbf{56}$ of $G_{4}=E_{7\left( 7\right) }$, $U$-duality group of $%
\mathcal{N}=8$, $D=4$ supergravity, is especially relevant. In this
case, the primitive tensors of the $\mathbf{56}$ of $E_{7\left(
7\right) }$ are related to the classification of discrete
$E_{7\left( 7\right) }\left( \mathbb{Z}\right) $-invariants; indeed,
as discussed \textit{e.g.} in \cite {Krut-1,Sen-1,Sen-2,Duff-1} (and
in particular in Sects. 3 and 4 of the fourth Ref. of \cite
{N=8-UV-finiteness}, and in App. E of \cite{Duff-2}), the discrete
$E_{7\left( 7\right) }\left( \mathbb{Z}\right) $-invariants are
given by the \textit{greatest common divisor} (\textit{gcd}) of
certain sets of numbers which correspond to covariant tensors of
$E_{7\left( 7\right) }\left( \mathbb{R}\right) $. Physically,
$E_{7\left( 7\right) }\left( \mathbb{Z}\right) $-invariants would
determine the algebraic classification of the charge orbits of
extremal black holes in presence of Dirac-Zwanziger-Schwinger
quantization
conditions. The currently known set of invariants of the $\mathbf{56}$ of $%
E_{7\left( 7\right) }\left( \mathbb{Z}\right) $ is given by the
\textit{gcd} of suitable projections of contractions of the
$K$-tensor itself with some
charge vectors $Q$'s \cite{Krut-1,Duff-1,Duff-2} (a manifestly $%
\left( SL\left( 2,\mathbb{R}\right) \times SO\left( 6,6\right)
\right) $-covariant formalism is worked out in \cite{Sen-1,Sen-2}).
Unfortunately, with the exception of the so-called
\textit{projective} black holes \cite{Krut-1}, the known set of
discrete invariants does not allow for a complete classification of
black hole states. Thus, it is natural to ask if the missing
invariants derived obtained by taking the \textit{gcd} of some independent tensors of the $%
\mathbf{56}$ of $E_{7\left( 7\right) }\left( \mathbb{R}\right) $,
given by
suitable tensor products of the $K$-tensor, suitably projected onto $E_{7}$%
-irreprs. and contracted with charge vectors $Q$'s. The
\textit{``master''} identity (\ref{Master-d=4}), yielding to various
relations constraining invariant structures of $E_{7}$, may actually
provide a systematic way to figure out a complete set of independent
invariant tensor structures.

Furthermore, the \textit{``master''} identity (\ref{Master-d=4}) is
relevant to derive and study the algebraic independence of
higher-order $U$-invariant polynomials appearing in the study of
multi-center black hole solutions \cite{FMOSY-1,irred-1}.

 We leave these interesting issues for future investigation.

\appendix

\section*{Acknowledgments}

A. M. and E. O. would like to thank the Department of Mathematics,
King's College, London UK, where part of this work was done, for
warm hospitality and inspiring environment. F.R. would like to thank
the Politecnico di Torino and Turin University for hospitality. A.
M. would like to thank Sergio Ferrara for enlightening discussions,
and E. O. would like to thank Mario Trigiante for useful
discussions. The work of A. M. has been supported in part by an INFN
visiting Theoretical Fellowship at SITP, Stanford University,
Stanford, CA, USA. The work of E. O. has been supported by the ERC
Advanced Grant no.226455, \textit{``Supersymmetry, Quantum Gravity
and Gauge Fields'' (SUPERFIELDS)}. The work of F.R. has been
supported by the STFC rolling grant ST/G000/395/1.

\section{\label{SU(2)-Convs}Conventions for $SU(2)$}

In this Appendix we summarise our conventions for $SU(2)$. The
generators are anti-Hermitian:
\begin{equation}
D_{i,a}{}^{b}=\frac{i}{2}\sigma _{i,a}{}^{b},
\end{equation}
where the $\sigma $'s denote the Pauli matrices. We are using a
negative-definite Cartan-Killing metric
\begin{equation}
g_{ij}=-\delta _{ij},
\end{equation}
which means that $D^{i}=-\frac{i}{2}\sigma _{i}$. The symmetric part
of the general Fierz identity reads
\begin{equation}
D_{b}^{i}{}^{c}D_{c}^{j}{}^{a}+D_{b}^{j}{}^{c}D_{c}^{i}{}^{a}={\frac{1}{2}}%
g^{ij}\delta _{b}^{a},  \label{symmetricproductsl2}
\end{equation}
and its contraction with $g_{ij}$ yields
\begin{equation}
D_{i}^{ab}D^{i,cd}=-{\frac{1}{4}}[\epsilon ^{ac}\epsilon
^{bd}+\epsilon ^{ad}\epsilon ^{bc}].  \label{symmm-1}
\end{equation}
From identities (\ref{symmetricproductsl2})-(\ref{symmm-1}), by
further contracting with $g_{ij}$ and/or $\epsilon _{ab}$, one can
obtain the following identities:
\begin{eqnarray}
D_{i,a}{}^{c}D_{c}^{i}{}^{b} &=&\frac{3}{4}\delta _{a}^{b}; \\
D_{a}^{i}{}^{b}D_{b}^{j}{}^{a} &=&\frac{1}{2}g^{ij}; \\
\lbrack D_{i},D_{j}]
&=&D_{i|a}{}^{c}D_{j|cb}-D_{j|a}{}^{c}D_{i|cb}=\epsilon
_{ij}{}^{k}D_{k}.
\end{eqnarray}
Finally, consistent with the negative definiteness of the metric,
the following product of Levi-Civita symbols is used:
\begin{equation}
\epsilon _{ikl}\epsilon _{j}{}^{kl}=-2g_{ij}\,.  \label{epseps}
\end{equation}

\section{\label{SO(10)}$E_{6}\supset SO(10)\times U(1)$}

For completeness, in the present Appendix we consider the reduction of some $%
E_{6}$-identities derived in Sect. \ref{E7--->E6}, according to the
group embedding (\ref{E6-emb}). The fundamental and the adjoint
irreprs. of $E_{6}$ respectively decompose as follows:
\begin{equation}
\mathbf{27}\rightarrow
\mathbf{10}_{-2}+\mathbf{16}_{1}+\mathbf{1}_{4}\quad
,\qquad \mathbf{78}\rightarrow \mathbf{45}_{0}+\mathbf{16}_{-3}+\mathbf{%
\overline{16}}_{3}+\mathbf{\overline{1}}_{0}\ ,
\label{E6toSO10decomp}
\end{equation}
where subscripts denote the $U\left( 1\right) $-charges. We here
denote with indices $A,B,...$ the vector $\mathbf{10}$ of $SO(10)$,
while the spinor
representations are denoted with $a,b,...$, where a lower index denotes the $%
\mathbf{16}$ and an upper index denotes the
$\mathbf{\overline{16}}$. Thus, the decomposition
(\ref{E6toSO10decomp}) implies the indices split as follows:
\begin{eqnarray}
V_{\widehat{M}} &\rightarrow &\left( V_{A}\ ,\ \ V_{a}\ ,\ \
V_{1}\right) ;
\\
V^{\widehat{M}} &\rightarrow &\left( V_{A}\ ,\ \ V^{a}\ ,\ \ V_{\overline{1}%
}\right) ; \\
V_{\widehat{\alpha }} &\rightarrow &\left( V_{AB}\ ,\ \ V_{a}\ ,\ \
V^{a}\ ,\ \ V_{1}\right) .
\end{eqnarray}

Before proceeding with the reduction, let us summarise our $SO(10)$
conventions. The charge conjugation matrix $C$ converts an upper $a$
index to a lower index $\dot{a}$, and viceversa:
\begin{equation}
C=\left(
\begin{array}{cc}
0 & C^{a\dot{b}} \\
C^{\dot{a}b} & 0
\end{array}
\right) ;
\end{equation}
it is antisymmetric and unitary, that is
\begin{equation}
C^{a\dot{b}}=-C^{\dot{b}a},
\end{equation}
and (``$\dagger $'' denotes Hermitian conjugation)
\begin{equation}
C_{a\dot{b}}^{\dagger }C^{\dot{b}c}=\delta _{a}^{c},\qquad C_{\dot{a}{b}%
}^{\dagger }C^{{b}\dot{c}}=\delta _{\dot{a}}^{\dot{c}}.
\end{equation}

The $\Gamma $-matrices have the form
\begin{equation}
\Gamma _{A}=\left(
\begin{array}{cc}
0 & \Gamma _{A,a}{}^{\dot{b}} \\
\Gamma _{A,\dot{a}}{}^{b} & 0
\end{array}
\right) ,
\end{equation}
and they satisfy the Clifford algebra
\begin{equation}
\{\Gamma _{A},\Gamma _{B}\}=2\eta _{AB},
\end{equation}
where $\eta _{AB}$ is the $D=10$ Minkowski metric (in the compact
case $\eta _{AB}=\delta _{AB}$), as well as the property
\begin{equation}
C\Gamma _{A}C^{\dagger }=-\Gamma _{A}^{T}.
\end{equation}
Note also that the matrix
\begin{equation}
(C\Gamma _{A})^{ab}=C^{a\dot{a}}\Gamma _{\dot{a}}{}^{b}
\end{equation}
is symmetric in the indices $ab$.

We now perform the reduction of some $E_{6}$-identities. The
Cartan-Killing metric of $E_{6}$ decomposes according to
\begin{equation}
g_{\widehat{\alpha }\widehat{\beta }}\rightarrow \left( -\delta _{AB}^{CD}\ (%
\mathbf{45\ 45})\ ,\ \ \delta _{a}^{b}\ (\mathbf{16\
{\overline{16}}})\ ,\ \ 1\ (\mathbf{1\ 1})\right) ,
\label{reductiongE6}
\end{equation}
whereas the invariant tensor $\delta _{\widehat{M}}^{\widehat{N}}$
branches as
\begin{equation}
\delta _{\widehat{M}}^{\widehat{N}}\rightarrow \left( \delta _{A}^{B}\ (%
\mathbf{10\ {10}}),\ \ \delta _{a}^{b}\ (\mathbf{16\
{\overline{16}}}),\ \ 1\ (\mathbf{1}\ \mathbf{\overline{1}})\right)
.  \label{reductiondeltaE6}
\end{equation}
Here $\delta _{AB}^{CD}\equiv \tfrac{1}{2}(\delta _{A}^{C}\delta
_{B}^{D}-\delta _{A}^{D}\delta _{B}^{C})$, and the minus sign in the
first term in the right-hand side of (\ref{reductiongE6}) has been
chosen for convenience, so that all coefficients in the reductions
under consideration are real. Note that all terms occurring in the
reduction must be invariant
tensors of $SO(10)\times U(1)$, and thus they trivially have vanishing $U(1)$%
-charge.

We consider the reduction of (\ref{ass-1}),
(\ref{ttdeltadeltaddE6}), (\ref {twos}) and (\ref{invconditionofd}),
as well as of the identities (\ref {ourmetric}) and
(\ref{gDDisdoverdlambda}) for $E_{6}$. For simplicity's sake, we do
not consider here the reduction of $E_{6}$-identities involving more
than three $E_{6}$-invariant tensors, as well as of identities
involving the $E_{6}$ structure constants. The reduction of the
invariant
tensors $d_{\widehat{M}\widehat{N}\widehat{P}}$ and $d^{\widehat{M}\widehat{N%
}\widehat{P}}$ reads
\begin{eqnarray}
d_{\widehat{M}\widehat{N}\widehat{P}} &\rightarrow &\left( \tfrac{1}{\sqrt{10%
}}\ \eta _{AB}\ (\mathbf{10\ 10\ 1})\ ,\ \ \tfrac{1}{2\sqrt{5}}\
(\Gamma
_{A}C^{\dagger })_{ab}\ (\mathbf{10\ 16\ 16})\right) ;  \notag \\
d^{\widehat{M}\widehat{N}\widehat{P}} &\rightarrow &\left( \tfrac{1}{\sqrt{10%
}}\ \eta ^{AB}\ (\mathbf{10\ 10\ \overline{1}})\ ,\ \
\tfrac{1}{2\sqrt{5}}\ (C\Gamma ^{A})^{ab}\ (\mathbf{10\
\overline{16}\ \overline{16}})\right) \ , \label{reductiondE6}
\end{eqnarray}
while the reduction of the generators $t_{\widehat{\alpha }\mid \widehat{M}%
}{}^{\widehat{N}}$ is
\begin{eqnarray}
t_{\widehat{\alpha }\mid \widehat{M}}{}^{\widehat{N}} &\rightarrow
&\left(
\begin{array}{l}
\tfrac{1}{4\sqrt{3}}\ (\Gamma _{AB})_{a}{}^{b}\ (\mathbf{45\ 16\ {\overline{%
16}}})\ ,\ \  \\
\\
\tfrac{1}{\sqrt{3}}\ \delta _{AB}^{CD}\ (\mathbf{45\ 10\ 10})\ , \\
\\
\tfrac{1}{3\sqrt{2}}\ \delta _{A}^{B}\ (\mathbf{1\ 10\ {10}})\ ,\  \\
\\
-\tfrac{1}{6\sqrt{2}}\ \delta _{a}^{b}\ (\mathbf{1\ 16\
\overline{10}})\ ,\
\\
\\
-\tfrac{\sqrt{2}}{3}\ (\mathbf{1\ 1\ \overline{1}})\ ,\  \\
\\
\tfrac{1}{2\sqrt{3}}\ (\Gamma _{A}C^{\dagger })_{ab}\ (\mathbf{16\
16\ 10})\ ,\ \ \tfrac{1}{2\sqrt{3}}\ (C\Gamma _{A})^{ab}\
(\mathbf{{\overline{16}}\
10\ {\overline{16}}})\ , \\
\\
-\tfrac{1}{\sqrt{6}}\ \delta _{a}^{b}\ (\mathbf{16\ 1\
\overline{16}})\ ,\ \ -\tfrac{1}{\sqrt{6}}\ \delta _{b}^{a}\
(\mathbf{{\overline{16}}\ 1\ {16}})\ ,
\end{array}
\right) ,  \notag \\
&&  \label{reductiontE6}
\end{eqnarray}
where $\Gamma _{AB}\equiv \Gamma _{\lbrack A}\Gamma _{B]}$.

One can then show that the reduction of $E_{6}$-identities (\ref{ass-1}), (%
\ref{ourmetric}), (\ref{gDDisdoverdlambda}) and (\ref{twos}) leads
to the
relations defining the Clifford algebra, together with trivial $\Gamma $%
-matrices identities. The reduction of (4.20) leads, among other
more trivial identities, to the well known $SO(10)$ Fierz identity
\begin{equation}
(C\Gamma ^{A})^{a(b}(C\Gamma _{A})^{cd)}=0,
\end{equation}
while the reduction of (4.16) leads, among the rest, to the Fierz
identity
\begin{equation}
(\Gamma _{A}C^{\dagger })_{ac}(C\Gamma ^{A})^{bd}-2\delta
_{a}^{d}\delta _{c}^{b}-\frac{1}{2}\delta _{a}^{b}\delta
_{c}^{d}-\frac{1}{4}(\Gamma _{AB})_{a}{}^{b}(\Gamma
^{AB})_{c}{}^{d}=0.
\end{equation}
Similarly, the reduction of the other $E_{6}$-identities derived in
Sect. \ref{E7--->E6} will give rise to additional $SO(10)$ $\Gamma
$-matrices identities, including additional Fierz identities.

\section{\label{Decomps}A Useful Decomposition}

A useful decomposition used in Sect. \ref{K-Tensor}, holding
\textit{at least} for all $U$-duality Lie groups $G_{4}$ of $D=4$
supergravities reported in Table 1, reads
\begin{equation}
t_{\alpha |M}^{\phantom{\alpha M}N}t_{\beta |NQ}=-t_{\alpha |MP}t_{\beta |NQ}%
\mathbb{C}^{PN}=\frac{1}{f}g_{\alpha \beta }\mathbb{C}_{MQ}+\frac{1}{2}%
f_{\alpha \beta }^{\gamma }t_{\gamma |MQ}+S_{(\alpha \beta )[MQ]}\,,
\label{ttDec}
\end{equation}
where the $G_{4}$-invariant tensor $S_{(\alpha \beta )[MQ]}$ is such
that
\begin{equation}
S_{(\alpha \beta )[MQ]}k^{\alpha \beta }\equiv 0\quad ;\quad
S_{(\alpha \beta )[MQ]}\mathbb{C}^{MQ}\equiv 0\,.  \label{Sirr}
\end{equation}

It is here worth pointing out that the left-hand side of
(\ref{ttDec}),
namely $t_{\alpha \mid M}^{\phantom{\alpha M}N}t_{\beta \mid NQ}$, is a $%
G_{4}$-singlet (because $t_{~MN}^{\alpha }$ is a $G_{4}$-singlet
itself; see
Eq. (\ref{ASS})). Thus, due to its symmetry properties, $t_{\alpha \mid M}^{%
\phantom{\alpha M}N}t_{\beta \mid NQ}$ enjoys a decomposition into \textit{%
irreducible} $G_{4}$-invariants terms, \textit{antisymmetric} under
the simultaneous exchanges $M\leftrightarrow Q$ and $\alpha
\leftrightarrow
\beta $. In other words, the adjoint indices and symplectic indices of $%
t_{\alpha \mid M}^{\phantom{\alpha M}N}t_{\beta \mid NQ}$ must have
opposite symmetry properties.

For simplicity's sake, let us derive (\ref{ttDec})-(\ref{Sirr}) in a
particular case, namely for $G_{4}=E_{7}$ (the generalization is
straightforward). The case $G_{4}=E_{7}$ pertains both to magic octonionic $%
\mathcal{N}=2$ ($G_{4}=E_{7\left( -25\right) }$, $J_{3}^{\mathbb{O}}$%
-related) supergravity and to $\mathcal{N}=8$ maximal theory ($%
G_{4}=E_{7\left( 7\right) }$, $J_{3}^{\mathbb{O}_{s}}$-related). For
this group, it holds that
\begin{eqnarray}
g_{\alpha \beta } &=&\mathbf{1}\in \mathbf{133}\times _{s}\mathbf{133}; \\
f_{\alpha \beta }^{~~\gamma }t_{\gamma } &=&\mathbf{1}\in
\mathbf{133}\times
_{a}\mathbf{133}; \\
\mathbb{C}_{MN} &=&\mathbf{1}\in \mathbf{56}\times _{a}\mathbf{56}.
\end{eqnarray}
Starting from the tensor product of the $\mathbf{R}\left( E_{7}\right) =%
\mathbf{56}$ and $\mathbf{Adj}\left( E_{7}\right) =\mathbf{133}$
irreprs., it follows that:
\begin{eqnarray}
\mathbf{56}\times _{s}\mathbf{56} &=&\mathbf{133}+\mathbf{1463}; \\
\mathbf{56}\times _{a}\mathbf{56} &=&\mathbf{1}+\mathbf{1539}; \\
\mathbf{133}\times _{s}\mathbf{133} &=&\mathbf{1}+\mathbf{1539}+\mathbf{7371}%
; \\
\mathbf{133}\times _{a}\mathbf{133} &=&\mathbf{133}+\mathbf{8645}\,.
\end{eqnarray}
These considerations lead to a decomposition of $t_{\alpha \mid M}^{%
\phantom{\alpha M}N}t_{\beta \mid NQ}$ that can contain only three
terms. Namely:

\begin{itemize}
\item  two terms with symmetric adjoint indices ($a$, $b\in \mathbb{R}$):
\begin{eqnarray}
ag_{\alpha \beta }\mathbb{C}_{MN} &=&\mathbf{1}=\mathbf{1}\times \mathbf{%
1\in }\left( \mathbf{133}\times _{s}\mathbf{133}\right) \times
\left(
\mathbf{56}\times _{a}\mathbf{56}\right) \mathbf{;} \\
bS_{\left( \alpha \beta \right) \left[ MN\right] } &=&\mathbf{1}\in
\left(
\mathbf{1539}\times \mathbf{1539}\right) \in \left( \mathbf{133}\times _{s}%
\mathbf{133}\right) \times \left( \mathbf{56}\times
_{a}\mathbf{56}\right) .
\end{eqnarray}

Notice that no other possibilities with symmetric adjoint indices
arise,
because\linebreak\ $\mathbf{1}\notin \left( \mathbf{7371}\times \mathbf{1539}%
\right) $.
\end{itemize}

\begin{itemize}
\item  one term with antisymmetric adjoint indices ($c\in \mathbb{R}$):

\begin{equation}
cf_{\alpha \beta }^{~~\gamma }t_{\gamma \mid \left( MN\right) }=\mathbf{1}%
\in \left( \mathbf{133}\times _{s}\mathbf{133}\right) \in \left( \mathbf{133}%
\times _{a}\mathbf{133}\right) \times \left( \mathbf{56}\times _{s}\mathbf{56%
}\right) .
\end{equation}
No other possibilities with antisymmetric adjoint indices arise, because $%
\mathbf{1}\notin \left( \mathbf{1463}\times \mathbf{133}\right) $, $\mathbf{1%
}\notin \left( \mathbf{1463}\times \mathbf{8645}\right) $, and $\mathbf{1}%
\notin \left( \mathbf{133}\times \mathbf{8645}\right) $.
\end{itemize}

Thus, $t_{\alpha \mid M}^{~~~Q}t_{\beta \mid QN}$ can be
$E_{7}$-irreducibly decomposed as follows:
\begin{equation}
t_{\alpha \mid M}^{~~Q}t_{\beta \mid QN}=ag_{\alpha \beta }\mathbb{C}%
_{MN}+bS_{\left( \alpha \beta \right) \left[ MN\right] }+cf_{\alpha
\beta }^{~~\gamma }t_{\gamma \mid \left( MN\right) }.
\label{decomp}
\end{equation}
In order to compute the constants $a,b,c$ $\in \mathbb{R}$, we
recall that
all terms of (\ref{decomp}) are \textit{irreducible}, as also implied by (%
\ref{Sirr}). Thus, by saturating (\ref{decomp}) with
$\mathbb{C}^{MN}$, one obtains
\begin{equation}
t_{\alpha \mid M}^{~~~Q}t_{\beta \mid QN}\mathbb{C}^{MN}=ag_{\alpha \beta }%
\mathbb{C}_{MN}\mathbb{C}^{MN}=fag_{\alpha \beta }\Leftrightarrow a=\frac{1}{%
f}\,.
\end{equation}
On the other hand, by recalling the definition (\ref{tt=ft}) of the
structure constants of the Lie algebra $\frak{g}_{4}$ of $G_{4}$ and using (%
\ref{decomp}), it follows that:
\begin{equation}
f_{\alpha \beta }^{~~\gamma }t_{\gamma \mid \left( MN\right) }\equiv
2t_{ \left[ \alpha \right| \mid M}^{~~~Q}t_{\left| \beta \right]
\mid QN}=t_{\alpha \mid M}^{~~~Q}t_{\beta \mid QN}-t_{\beta \mid
M}^{~~~Q}t_{\alpha \mid QN}=2cf_{\alpha \beta }^{~~\gamma }t_{\gamma
\mid \left( MN\right) }\Leftrightarrow c=\frac{1}{2}.
\end{equation}
Obviously, the constant $b$ can be reabsorbed in a re-definition of $%
S_{\left( \alpha \beta \right) \left[ MN\right] }$. Thus, the
irreducible decomposition (\ref{ttDec}) has been proved to hold.
$\blacksquare $

\end{document}